\documentclass[aps,prx,twocolumn,superscriptaddress]{revtex4-2}

\usepackage{standalone}

\usepackage{amsthm}
\usepackage{lipsum}

\usepackage{amsfonts, amsmath, amssymb}
\usepackage{graphicx}
\usepackage{dcolumn}
\usepackage{bm}
\usepackage{bbm}
\usepackage{MnSymbol}
\usepackage{dsfont}
\usepackage{MnSymbol}
\usepackage[normalem]{ulem}
\usepackage{color}
\usepackage[utf8]{inputenc}
\usepackage[colorlinks,citecolor=blue,linkcolor=blue,urlcolor=blue]{hyperref}
\usepackage{tikz}
\usepackage{braket}
\usepackage{physics}
\usepackage{color}
\usepackage{soul}
\usepackage{mathtools}
\usepackage{float}
\usepackage{esvect}
\usepackage{subfigure}
\usepackage{stmaryrd} 
\usepackage{enumerate}
\usepackage{bbold}
\usepackage{appendix}
\usepackage{courier}
\newcommand{\N}{\mathcal{N}}
\newcommand{\cri}[1]{{\color{black}{#1}}} 
\newcommand{\fra}[1]{{\color{black}{#1}}} 
\usepackage{physics}

\usepackage{xcolor}
\newcommand{\jac}[1]{{\textcolor{black}{#1}}} 
\include{physics}

\makeatletter
\def\maketitle{
	\@author@finish
	\title@column\titleblock@produce
	\suppressfloats[t]}
\makeatother

\begin{document}

\begin{abstract}
\noindent 
We introduce a variational method for simulating the dynamics of interacting open quantum bosonic systems deep in the quantum regime. The method is based on a multi-dimensional Wigner phase-space representation and employs a Variational Multi-Gaussian (VMG) ansatz, whose accuracy is systematically controlled by the number of Gaussian components. The variational equations of motion are derived from the Dirac–Frenkel principle and evaluated efficiently by combining the analytical structure of Gaussian functions with automatic differentiation. As a key first physical application, we study a driven-dissipative two-dimensional Bose–Hubbard lattice with two-boson coherent driving and two-body losses. Using our dynamical approach, we compute the finite-size scaling of the Liouvillian spectral gap—extracted from the relaxation dynamics—which vanishes in the thermodynamic limit. Our results reveal critical slowing down with dynamical exponents of the 2D quantum Ising universality class, demonstrating the power of our method to capture complex quantum dynamics in large open systems. 
\end{abstract}

\title{Variational Multi-Gaussian Phase-Space Bosonic Dynamics
\\ via Automatic Differentiation}
\author{Jacopo Tosca}
\affiliation{Universit\'{e} Paris Cit\'e, CNRS, Mat\'{e}riaux et Ph\'{e}nom\`{e}nes Quantiques, 75013 Paris, France}
\author{Francesco Carnazza}
\affiliation{Universit\'{e} Paris Cit\'e, CNRS, Mat\'{e}riaux et Ph\'{e}nom\`{e}nes Quantiques, 75013 Paris, France}
\author{Luca Giacomelli}
\affiliation{Universit\'{e} Paris Cit\'e, CNRS, Mat\'{e}riaux et Ph\'{e}nom\`{e}nes Quantiques, 75013 Paris, France}
\author{ Cristiano Ciuti}
\affiliation{Universit\'{e} Paris Cit\'e, CNRS, Mat\'{e}riaux et Ph\'{e}nom\`{e}nes Quantiques, 75013 Paris, France}

\date{\today}
\maketitle
\cri{Complex quantum systems, consisting of many interacting degrees of freedom evolving under coherent dynamics, dissipation, and external driving, represent one of the central frontiers of modern condensed matter physics \cite{girvin2019modern}. Their rich behavior underlies a wide range of phenomena, from emergent collective phases to nonequilibrium criticality, and forms the foundation of rapidly developing quantum technologies. Insights gained from studying such high dimensional quantum dynamics are also valuable for the analysis of complex classical systems that involve similarly large configuration spaces. Developing theoretical frameworks capable of addressing this level of complexity remains a major challenge at the interface between quantum and classical many body science. 

Within this broad context, interacting open quantum bosonic systems play a central role in modern quantum science. Photonic platforms provide a clear illustration, since nonlinear media allow photons to acquire effective interactions and collective behavior that give rise to nonequilibrium quantum phases~\cite{Carusotto2013,Ozawa2019,Carusotto2020photonic}. The same nonlinear mechanisms are central to the development of new architectures for quantum information processing, where engineered systems support the deterministic creation and manipulation of photonic Schrödinger's cat states~\cite{Mirrahimi2014dynamically,Leghtas2015,Weigand2018,Guillaud2019,Grimm2020kerrcat,Larsen2021deterministic,joshi2021quantum,Gautier2022,Chamberland2022,Xiaozhou2023,Hajr2024,Marquet2024,ding2025quantum,Hutin2025,Ming2025}. These platforms combine strong interactions, dissipation, and large numbers of modes, and therefore exemplify the need for theoretical approaches capable of treating complex dynamics beyond standard approximations. Describing the dynamics of open quantum bosonic systems with a large number of modes remains highly nontrivial.}

The full quantum dynamics of interacting bosons—especially under non-equilibrium and dissipative conditions \cite{RMP2025}—poses severe challenges due to the infinite-dimensional Hilbert space of each mode, the presence of strong quantum correlations, and the need to capture both transient and steady-state behavior. Powerful methods, such as those based on the Density Matrix Renormalization Group \cite{schollwock2011density}, Matrix Product States, and Tensor Networks~\cite{orus2019tensor}, have been applied to driven-dissipative settings mostly for spin lattices~\cite{Kilda2021,Hryniuk2024}, while open bosonic systems, especially in more than one dimensions, remain a more difficult hurdle. \\
\cri{The emergence of machine learning approaches has sparked interest in Neural Quantum States~\cite{carleo2017solving}. These methods provide expressive variational \textit{ansätze} for quantum many-body wavefunctions and have reached state-of-the-art performance in simulating ground states of lattice spin models and related systems~\cite{Carrasquilla2021, Vicentini2019, Hartmann2019, PhysRevB.99.214306, Nagy2019, Torlai2018NNQST, Carnazza_2022, Luo2022,Denis2025}. However, extending them to bosonic statistics or continuous variables in driven-dissipative regimes has proven to be difficult.} \cri{As emphasized in recent work on bosons~\cite{Denis2025}, “accurate neural representations of the density matrix are still unavailable’’ for such settings.} Other techniques use low-rank representations of the many-body density matrix \cite{Finazzi2015Corner,Santos2025} and can reach accuracies comparable to exact diagonalization even with bosons. \cri{In particular, the Corner Space Renormalization approach has proven effective in describing non-equilibrium quantum phase transitions~\cite{rota2019quantum}, yet it is constrained to steady-state properties and to systems with moderately low von Neumann entropy, which limits the size of the systems that can be simulated.}

For bosonic systems, a natural representation of the many-body density matrix is provided by the Wigner function \cite{polkovnikov2010phase}, which maps the quantum state to a quasi-probability distribution in phase space, with each bosonic mode described by a complex variable (or a pair of real quadratures). The time evolution of the Wigner function, governed by the Lindblad master equation, takes the form of a partial differential equation with higher-order derivatives with respect to the phase-space bosonic variables. 
\jac{Stochastic methods based on phase-space representations are highly effective in many scenarios; however, they often suffer from convergence problems and stability issues in regimes characterized by strong quantum correlations~\cite{Deuar2019, Li2025}.}

To date, no approach has succeeded in providing a scalable and accurate variational method in the fully quantum regime based on a phase-space representation. Nevertheless, a phase-space representation, being functional in nature, opens the door to using well-defined sets of analytical functions—potentially enabling exact manipulations even in regimes with strong negativities and correlations. \cri{In particular, given the structural similarity between the Lindblad equation in phase space and a high-dimensional partial differential equation, one may exploit analytical function families to construct compact yet highly expressive variational ansätze.} In principle, this could be achieved with automatic differentiation techniques \cite{Wengert1964, Margossian2019}, a \cri{powerful} and recent technique developed in computer science, which fully exploits modern computational parallel \cri{architectures} such as those based on Graphic Processing Units (GPUs).

\begin{figure*}[t!]
  \centering
  \includegraphics[width=1\textwidth]{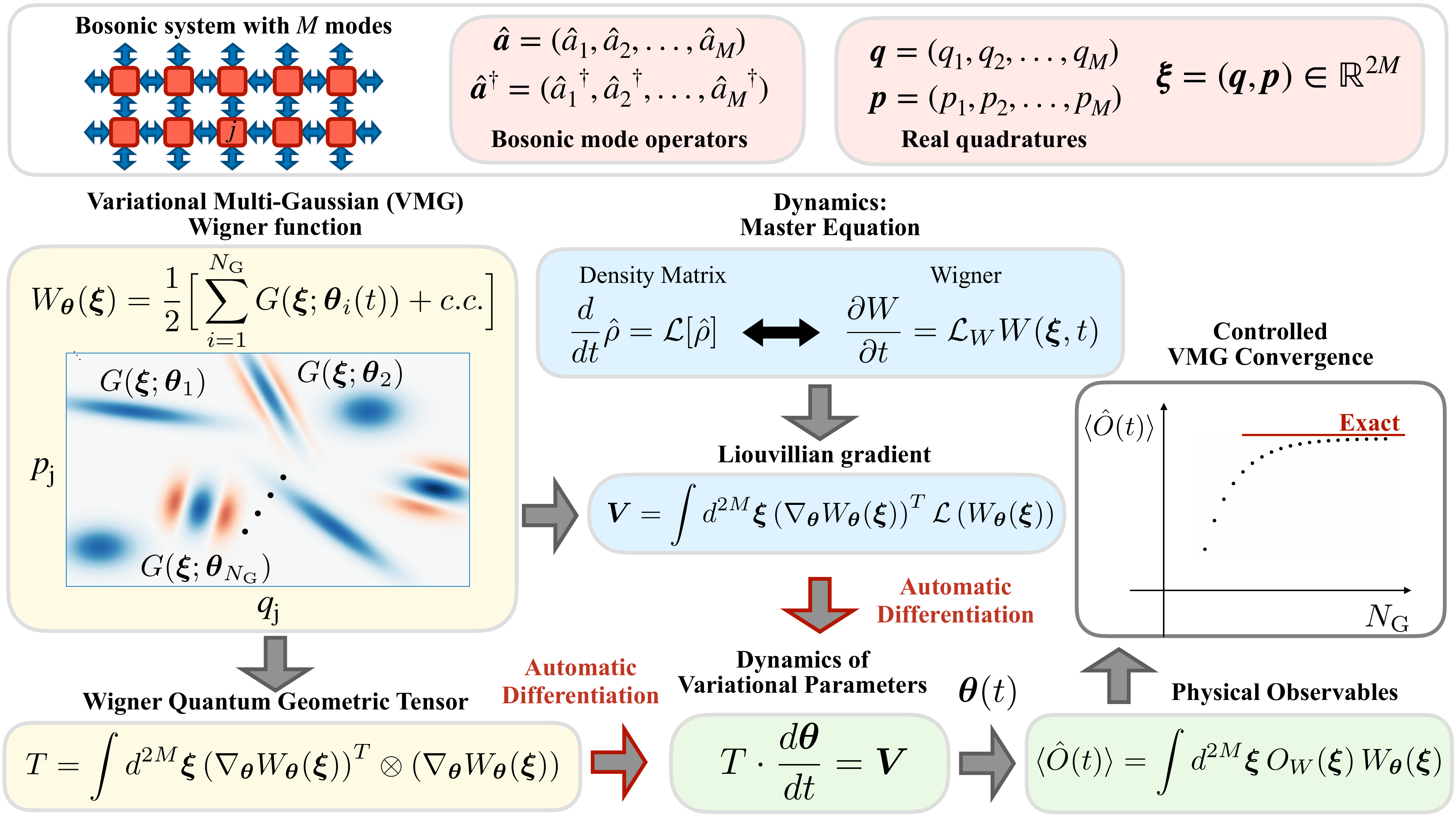}
  \caption{Scheme of the variational method demonstrated in this work to calculate accurately the dynamics of a quantum system consisting of $M$ bosonic modes.   
  The ansatz for the Wigner function $W_{\bm \theta }$, representing the many-boson density matrix $\hat \rho$, is expressed as a sum of $N_{\rm G}$ Gaussian functions (and their complex conjugates $c.c.$)  each characterized by its variational parameters $\bm \theta_i$: a normalization factor, the center of the Gaussian and its covariance matrix. 
 The Wigner function is defined with respect to the variable $\boldsymbol{\xi} = (\boldsymbol{q},\boldsymbol{p})$ where the vectors $\boldsymbol{q} = (q_1,q_2, ..., q_{M})$ and $\boldsymbol{p} = (p_1,p_2, ..., p_{M})$ contain the real quadrature variables describing the $M$ bosonic modes. 
Since the centers can be complex, the resulting (squeezed) Gaussians are modulated by sinusoidal functions and can therefore take both positive (shown in blue) and negative (shown in red) values.
The time evolution of the  variational parameter vector $\bm \theta$ is governed by the Lindblad master equation (Eq.~\eqref{eq:Lindblad}) through the Dirac-Frenkel variational principle. Specifically, the time derivative $d\boldsymbol{\theta}/dt$ of the variational parameters depends on the Wigner-space quantum geometric tensor $T$ and the Liouvillian gradient vector $\boldsymbol{V}$. 
Both quantities are determined by multidimensional phase-space integrals, which can be computed exactly due to the analytical properties of Gaussian functions and efficiently using automatic differentiation techniques.
Any time-dependent observable $\hat O(t)$ can be evaluated efficiently via the analytical integral of its Weyl symbol $O_W$ over the Wigner function at time $t$. The VMG ansatz ensures controlled convergence toward the exact solution by increasing the number $N_{\rm G}$ of Gaussian functions.}
  \label{fig:scheme}
\end{figure*}

In this work, we introduce a novel variational phase-space method for open quantum bosonic systems, based on an analytical Variational Multi-Gaussian (VMG) ansatz for the Wigner function (see Fig.~\ref{fig:scheme}).  Our approach achieves accuracies comparable to exact diagonalization while maintaining exceptional scalability. The ansatz consists of a sum of Gaussian components, whose number can be systematically increased to control the precision of the solution. The time evolution of the variational parameters is derived analytically and exactly using the Dirac–Frenkel variational principle, exploiting the phase-space functional representation, the analytical properties of multi-Gaussian functions and fully leveraging automatic differentiation~\cite{griewank2008evaluating,Margossian2019}. \cri{The structure of the approach also suggests potential applicability to classical dynamical problems that involve partial differential equations in high-dimensional phase spaces \cite{Cho2016PDF,   Han2018,Achdou2007PDEFinance}.}

This paper is organized as follows. In Section~\ref{sec:open}, we outline the phase-space framework used to describe open quantum bosonic systems.
Section~\ref{sec:variational_ansatz} introduces the Variational Multi-Gaussian (VMG) ansatz, and in Section~\ref{sec:Dynamics_var} we derive the equations for the time-dependent variational parameters from the Dirac–Frenkel variational principle.
In Section~\ref{sec:Generalized Gaussian moments}, \cri{we discuss the implementation of the method using automatic differentiation}.
In Section~\ref{sec:single-kerr}, we benchmark the method on a single driven dissipative Kerr quantum parametric oscillator by comparing both the time evolution of observables and the Wigner function to exact results. \cri{In Section \ref{sec:Kerr-nonlocal}, we benchmark our method to a system consisting of three Kerr parametric oscillators coupled via nonlocal quantum losses.}
In Section~\ref{sec:bose-hubbard-steady}, we apply the approach to a two-dimensional driven dissipative Bose–Hubbard lattice with two-boson coherent driving and losses, focusing on steady-state properties. \cri{We benchmark our results with steady-state data available in the literature.}
In Section~\ref{sec:bose-hubbard-dynamics}, we examine the dynamical behavior of the same system and \cri{reveal for the first time the associated dynamical quantum critical exponents}.
Conclusions and perspectives are presented in Section~\ref{conclusions}.
\section{Open quantum dynamics in phase space}
\label{sec:open}

The dynamics of the density matrix $\hat \rho$ of an open quantum bosonic system, in the presence of weak coupling to a Markovian environment, is described by the Lindblad master equation \cite{Breuer2002, Fazio2025}: 
\begin{equation}\label{eq:Lindblad}
    \frac {\partial \hat \rho(t)} {\partial t} = \mathcal{L} \hat \rho(t) = - i [\hat H, \hat \rho(t) ] + 
    \sum_j \mathcal{D}_j[\hat \rho(t)].
\end{equation}
In this master equation, $\mathcal L$ stands for the Liouvillian superoperator which depends on the system's Hamiltonian $\hat H$ which, in turn, encodes the unitary evolution of the system. The dissipation superoperators $\mathcal { D}_j$ account instead for the interaction channels with the external environment, whose action on the density matrix $\hat \rho$ is fully characterized by the jump operators $\hat \Gamma_j$:

\begin{equation}\label{eq:dissipation}
    \mathcal{ D}_j {\hat \rho(t)}=
 {\hat \Gamma_{j}} {\hat \rho(t)} \hat \Gamma_{j}^{\dagger}-\frac{1}{2}
\left\{
\hat \Gamma_{j}^{\dagger} \hat \Gamma_{j}, 
{\hat \rho(t)}
\right\}.
\end{equation}
All terms in the Lindblad master equation in Eq.~\eqref{eq:Lindblad} can be expressed in terms of the bosonic annihilation and creation operators, $ \hat{\boldsymbol{a}} = (\hat{a}_1, \hat{a}_2, \ldots, \hat{a}_M) $ and $ \hat{\boldsymbol{a}}^\dagger = (\hat{a}_1^\dagger, \hat{a}_2^\dagger, \ldots, \hat{a}_M^\dagger) $, corresponding to the $ M $ modes of the system. These obey the standard
commutation relations ${\left[\hat{a}_j, \hat{a}_k^{\dagger}\right]=\delta_{j k},}$ and ${\left[\hat{a}_j, \hat{a}_k\right]=\left[\hat{a}_j^{\dagger}, \hat{a}_k^{\dagger}\right]=0 }$.

The representation of a quantum system as a density matrix $\hat \rho$ is not the unique option.
By analogy to classical statistical mechanics, one can show that the same information can be stored in a phase-space distribution \cite{Wigner1932}. 
In this framework, the density matrix $\hat \rho$ is represented by a normalized quasi-probability density function $W(\boldsymbol{q},\boldsymbol{p})$, known as the Wigner function, where
$\boldsymbol{q} = (q_1,q_2,...,q_M)$ and $\boldsymbol{p} = (p_1,p_2,...,p_M)$ are the phase-space real quadratures and $M$ is the total number of bosonic modes.
The designation $\textit{quasi}$-probability reflects that, in stark contrast to classical probability density functions, the Wigner function might assume negative values. Note that there are other possible valid phase-space representations of the density matrix, most notably the Glauber-Sudarshan P \cite{Glauber1963,Sudarshan1963} and the Husimi Q \cite{Husimi1940}, which will not be considered in this work.

The Wigner function on the phase-space variables $(\boldsymbol q,\boldsymbol p)$ is defined in terms of the density matrix  
$\hat \rho$ \cite{Wigner1932}: 
\begin{equation}\label{eq:Wigner}
W(\boldsymbol{q},\boldsymbol{p}) = 
\frac{1}{(\jac{2}\pi \hbar)^{M}} 
\int_{-\infty}^{+\infty} 
d \boldsymbol{y} \langle \boldsymbol{q} - \boldsymbol{\jac{\frac{y}{2}}} | \hat \rho | \boldsymbol{q} + \boldsymbol{\jac{\frac{y}{2}}} \rangle  e^{i \boldsymbol{p} \cdot \boldsymbol{y} / \hbar}. 
\end{equation}

Besides providing an alternative formulation of quantum mechanics, this phase-space formulation recasts quantum mechanics from its usual operator-based (algebraic) picture into a differential one. In this formulation, quantum operators acting on the density matrix are replaced by polynomials and derivatives, defined on the phase-space domain, which act on the Wigner function. 
The transformation acting on the quantum state $\hat \rho$ in Eq.~\eqref{eq:Wigner} also maps the ``algebraic'' Liouvillian operator $\mathcal{L}$ to a differential one, $\mathcal{L}_W$, whose structure is expressed in detail in the Appendix \ref{appendix:phase_space}. 
In phase-space, the dynamics of the density matrix in Eq.~\eqref{eq:Lindblad} can be recast as: 
\begin{equation}\label{eq:phase_space_lindblad}
\begin{aligned}
    \frac{\partial W(\boldsymbol{q},\boldsymbol{p},t)}{\partial t} &= \mathcal L_W W(t) = \\
    & \{H_W, W(t)\}_{MB} + \sum_j \gamma_j {\mathcal{D}_W}_j[W(t)],
\end{aligned}
\end{equation}
where $\{\circ , \circ \}_{MB}$ and $\mathcal D_W$ are, respectively, the Moyal brackets and the phase-space dissipation superoperator, both defined in detail in the Appendix \ref{appendix:phase_space}.

\section{Variational Multi-Gaussian ansatz for Wigner function}\label{sec:variational_ansatz}

In this work, we adopt a variational approach in which we consider an ansatz for the Wigner function consisting of a sum of $ N_{\rm G} $ Gaussian functions, namely:
\begin{equation}
    W_{\bm \theta}(\boldsymbol \xi) = \frac{1}{2}
    \left(
    \sum_{i=1}^{N_{\rm G}} G(\boldsymbol \xi; \boldsymbol\theta_i)+ c.c.
    \right) \equiv \sum_{i=1}^{N_{\rm G}} {\rm Re}[G(\boldsymbol \xi; \boldsymbol\theta_i)], 
    \label{eq:vmg}
\end{equation}
where $c.c.$ stands for the complex conjugate and $\boldsymbol{\theta} = (\boldsymbol{\theta}_1,\boldsymbol{\theta}_2, ...,  \boldsymbol{\theta}_{N_G}$) is the vector of variational parameters.
Each Gaussian function of the multi-mode phase-space variables $\boldsymbol{\xi} = (\boldsymbol{q},\boldsymbol{p})$  is defined as \cite{Bourassa2021, Walschaers2021}
\begin{equation}
\begin{aligned}
    &G(\boldsymbol \xi ; \boldsymbol{\theta_i})
    = \\
    &\frac{c_i}{\jac{(2\pi)^{M}}\sqrt{ \det(\Sigma_i)}} \exp\left[{-\frac{1}{2} (\boldsymbol\xi-\boldsymbol{\mu}_i)^\jac{T} \Sigma_i^{-1}(\boldsymbol\xi-\boldsymbol{\mu}_i) } \right],
\end{aligned}\label{eq:Gaussian}
\end{equation}
where  $\boldsymbol \theta_i = (c_i,\boldsymbol \mu_i,\Sigma_i)$ is the parameter vector including the variational parameters of the $i-$th Gaussian, namely the normalization coefficient $c_i$, the center $\boldsymbol \mu_i$ and the covariance matrix $\Sigma_i$. Note that the normalization also depends on  $M$, i.e. the  number of bosonic modes. \\
In order for an ensemble of Gaussians to be expressive enough to capture Wigner functions with negative fringes due to macroscopic quantum interference (such as those present for the so-called ``Schrödinger cat states'' \cite{Leghtas2013,Yurke1986}), we allow the centers $\boldsymbol \mu$ to take complex values: $\boldsymbol \mu =\boldsymbol \alpha + \mathrm{i} \boldsymbol \beta$, where $\mathrm{i}$ is the imaginary unit.
The real part of a complex-centered Gaussian corresponds in fact to a Gaussian envelope modulated by a sinusoidal function. This form has been shown to exactly reproduce the negative fringes of cat states \cite{Castagnos97, Bourassa2021, Vogel1989}.
\jac{The condition for the density matrix to have unitary trace ($\text{Tr}[\hat \rho] = 1$) requires a normalization condition over the coefficients $c_i$:}
\begin{equation}
    \sum_{i=1}^{N_{\rm G}} c_i =1.
\end{equation}
Moreover, in our ansatz we will require the Hermitian covariance matrices $\Sigma_i$ to be real and semi-definite positive. 
Note that there is not a unique way to parametrize the covariance matrices $\Sigma (\bm \theta)$ as functions of the variational parameters $\bm \theta$. We discuss in detail the choice for our variational parameters and their initialization in Appendix \ref{app:parametrization}. An alternative ansatz, based on linear superposition of Gaussian states for individual quantum trajectories, has also been recently proposed \cite{Qu2025}. Their stochastic method has been exploited to investigate dynamical effects for a Kondo model.  

The considered variational multi-Gaussian ansatz, reported in Eq.~\eqref{eq:vmg}, offers a controllable level of expressivity: the accuracy of the description for arbitrary observables can be systematically improved by increasing the number of Gaussian components. Moreover, the Gaussian structure enables an efficient and fully analytical derivation of the equations governing the time evolution of the variational parameters.\\
\jac{We compute the time evolution of the ansatz $W_{\bm \theta}$ using automatic differentiation~\cite{Bettencourt2019, jax2018}, a strategy enabled by the fact that the action of any operator on the Wigner function corresponds to a polynomial in the phase-space variables $\bm \xi$ and derivatives $\partial_{\bm \xi}$. As detailed below, these polynomials are evaluated via repeated partial differentiation, which is handled efficiently by the automatic differentiation algorithm.}

\jac{\section{Dynamical equations of the variational parameters}}
\label{sec:Dynamics_var}
\begin{figure*}
    \centering
        \includegraphics[width=\textwidth]{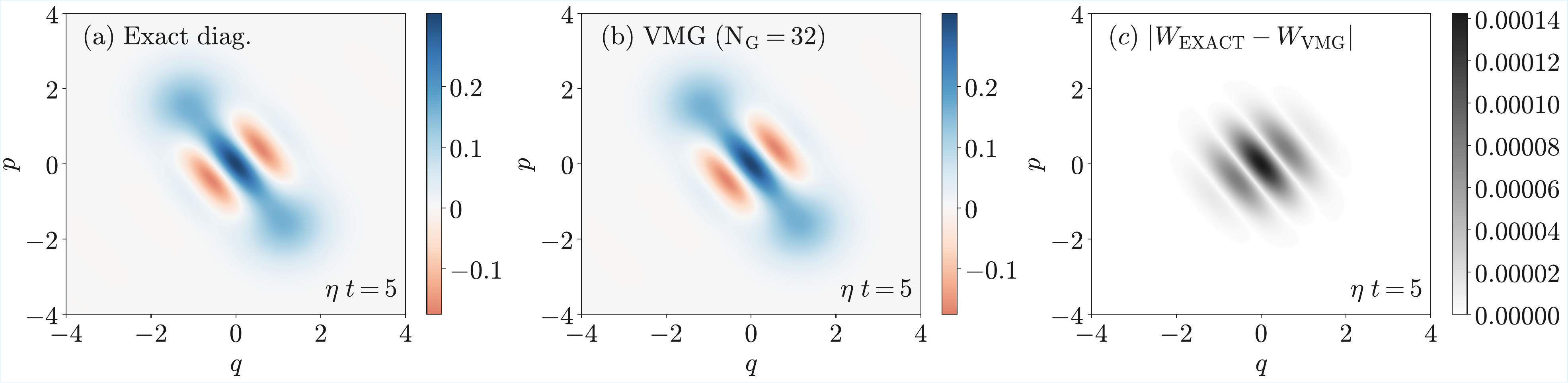}
    \caption{(a) Wigner function at time $ t = 5/\eta $ obtained via exact diagonalization techniques (specifically via Fock base expansion with bosonic cutoff $\Lambda_{\rm {b}} = 100$) for the single-mode dissipative Kerr parametric oscillator. The non-unitary dynamics is governed by the Hamiltonian described by Eq.(\eqref{eq:single-mode}) (boson-boson interaction with two-boson driving) and by two-boson loss rate $\eta$. \jac{Initial state for Fock evolution: vacuum state. Initial state for VMG is a small perturbation of the vacuum: $W_{\rm init,\, 0}$ (for further details see App.~\ref{app:parametrization}).}
(b) Wigner function at the same time computed using the Variational Multi-Gaussian (VMG) ansatz with $32$ complex Gaussians.
(c) Absolute value of the difference between the exact and the variational Wigner functions, highlighting the remarkable accuracy of the VMG ansatz, which, unlike the exact integration approach, is scalable to systems with many modes.
Other parameters: boson-boson interaction $ U/\eta = 1 $, single-boson loss-rate $\gamma = 0$, detuning $\Delta = 0$ and two-boson driving complex amplitude $ G/\eta = 3 + 3i $.
}
\label{cats}
\end{figure*}
In order to determine the dynamics of an open bosonic quantum system using the Variational Multi-Gaussian ansatz for the Wigner function, we apply the time-dependent Dirac–Frenkel variational principle \cite{Yuan2019, Lasser2022}. This yields the equations governing the evolution of the optimal variational parameters. Considering a variational ansatz $\hat{\rho}_{\boldsymbol{\theta}(t)}$ for the density matrix, the corresponding differential equations for the variational parameters read:

\begin{equation}
    T \cdot \frac{d}{dt}\boldsymbol \theta = \boldsymbol V, \label{eq:DiracFrank}
\end{equation}
where 
\begin{equation}
\begin{aligned}
    &T_{rs} = \textrm{Tr}
    [(\nabla_{\boldsymbol\theta} \hat{\rho}_{\boldsymbol{\theta}})_r^\dagger
    (\nabla_{\boldsymbol\theta} \hat{\rho}_{\boldsymbol{\theta}})_s
    ], \\
    & {V}_r = \textrm{Tr}
    [(\nabla_{\boldsymbol\theta} \hat \rho_{\boldsymbol{\theta}})_r^\dagger
    \mathcal{L}\hat \rho].
\end{aligned}
\end{equation}
Here, $T_{rs}$ is the quantum geometric tensor, which is solely determined by the structure of the variational ansatz.  The vector $V_r$ represents instead the Liouvillian gradient, which depends both on the structure of the variational ansatz and on the form of the master equation governing the system's dynamics.

In the Wigner representation 
these quantities can be expressed as phase-space integrals, namely:
\begin{equation}
\begin{aligned}
\label{T_matrix}
T & =
\int  d^{2M} \bm \xi
\Big( \nabla_{\boldsymbol\theta} W_{\boldsymbol\theta}(\boldsymbol\xi) \Big) ^T \otimes
\Big( \nabla_{\boldsymbol\theta} W_{\boldsymbol\theta}(\boldsymbol\xi) \Big)
, \\
\bm V & =\int 
d^{2M} \bm \xi
 \Big( \nabla_{\boldsymbol\theta} W_{\boldsymbol\theta}(\boldsymbol\xi) \Big) ^T
 \mathcal{L}_W (W_{\boldsymbol\theta}(\boldsymbol\xi)), 
\end{aligned}
\end{equation}
where $\boldsymbol d^{2M}\xi = \jac{(2\pi\hbar)^M}d^{M}\boldsymbol q \ d^{M}\boldsymbol p$. 
\jac{As we show in detail in the next section, the multi-Gaussian ansatz enables the analytical computation of both the Quantum Geometric Tensor ($T$) and the Liouvillian gradient ($\boldsymbol{V}$) for a broad class of interacting Hamiltonians.}\\

\section{Generalized Gaussian moments via automatic differentiation}\label{sec:Generalized Gaussian moments}
A central challenge for time-dependent variational methods is to design an ansatz that is both highly expressive and enables an efficient evaluation of the quantum geometric tensor and the Liouvillian gradient. As we demonstrate below, our variational approach based on automatic differentiation successfully meets both requirements and opens the door to remarkable applications that have remained out of reach for existing methods.

Quantum states in the Variational Multi-Gaussian form are especially amenable to analytical treatment for their variational dynamical evolution.
Specifically, for a VMG Wigner function $W_{\bm \theta}$ as in Eq.~\eqref{eq:vmg}, the Wigner quantum geometric tensor $T$ can be conveniently computed by taking the gradients $\nabla_{\boldsymbol \theta}$ outside the sign of the integral in Eq.~\eqref{T_matrix}. Namely, we get the expression:
\begin{equation}
\begin{aligned}
&T[\boldsymbol \theta] = \sum_{m,n = 1}^{N_{\rm G}}
\nabla_{\boldsymbol \theta} \nabla_{\boldsymbol \theta'} 
\int d^{2M} \bm \xi \times\\
 &\qquad
{\rm Re}[ G(\bm \xi;\boldsymbol \theta)]
{\rm Re}[ G(\bm \xi;\boldsymbol \theta')]
\Big|_{\substack{
\boldsymbol \theta = \boldsymbol \theta_m \\
\boldsymbol \theta' = \boldsymbol \theta_n}}.
\end{aligned}
\label{eq:T_tensor}
\end{equation}

In an analogous fashion, the Liouvillian gradient $\bm V$ in Eq.~\eqref{T_matrix} can be recast as:
\begin{equation}
\begin{aligned}
&V[\boldsymbol \theta] = \sum_{m,n = 1}^{N_{\rm G}}
\nabla_{\boldsymbol \theta} 
\int d^{2M} \bm \xi \times\\
&\qquad 
{\rm{Re}}[G(\bm \xi;\boldsymbol \theta)]\mathcal{L}_W
{\rm{Re}}[G(\bm \xi;\boldsymbol \theta')]
\Big|_{\substack{
\boldsymbol \theta = \boldsymbol \theta_m \\
\boldsymbol \theta' = \boldsymbol \theta_n}}.
\end{aligned}
\label{eq:V_theta}
\end{equation}
Here, the gradient $\nabla_{\bm \theta}$ can be taken outside of the integral because it applies only to the first term (${\rm{Re}}[G(\bm \xi;\boldsymbol \theta)]$), while it is zero on the second ($\mathcal L_W {\rm{Re}}[G(\bm \xi;\boldsymbol \theta')]$). \\
As detailed in Appendix \ref{Bose-hubbardPhase}, the Liouvillian operator $\mathcal{L}$ can be decomposed as $\mathcal{L}_W=\sum_\mathbf{I} \mathcal L_{W,\mathbf{I}}$, where each Liouvillian component is such that:
\begin{equation}
    \mathcal L_{W,\mathbf{I}} \propto 
    \xi_{i_1}...\xi_{i_l}
    \partial_{\xi_{j_1}}...\partial_{\xi_{j_k}}.
\end{equation}
This means that, in order to compute the Liouvillian gradient vector $\bm V$, we need to evaluate the {\it generalized Gaussian moments} defined as follows:
\begin{equation}
\begin{aligned}
    &\llangle
    \xi_{i_1}...\xi_{i_l} 
    \partial_{\xi_{j_1}}...\partial_{\xi_{j_k}}
    \rrangle_{\bm \theta,\bm \theta'}
   = \int d^{2 M} \bm \xi \times \\
    &\quad {\rm{Re}}[G(\bm \xi;\boldsymbol \theta)]
    \xi_{i_1}...\xi_{i_l}
    \partial_{\xi_{j_1}}...\partial_{\xi_{j_k}}
    {\rm{Re}}[G(\bm \xi;\boldsymbol \theta')].
    \label{eq:ggm}
\end{aligned}
\end{equation}
The Eq.~\eqref{eq:ggm}, for the generalized Gaussian moments, can be streamlined by recasting them as derivatives with respect to vanishing currents $\boldsymbol J$ and $ \tilde {\boldsymbol J}$ of a generating function $\mathcal{Z}[\bm J,\tilde {\bm J}]$:
\begin{equation}
\begin{aligned}
    &\llangle
    \xi_{i_1}...\xi_{i_l} 
    \partial_{\xi_{j_1}}...\partial_{\xi_{j_k}}
    \rrangle_{\boldsymbol \theta , \boldsymbol \theta'}
   =\\&
    \partial_{J_{i_1}}...\partial_{J_{i_l}}
    \partial_{\tilde J_{j_1}}...\partial_{\tilde J_{j_k}}
    \mathcal {Z}[\boldsymbol J, \tilde {\boldsymbol J}]|_{\boldsymbol J=0, \tilde {\boldsymbol J} =0} \,.
\end{aligned}
\label{eq:re_ggm}
\end{equation}
The functional form of the generating function $\mathcal Z[\boldsymbol J,\tilde {\boldsymbol J}]$ is reported in Appendix \ref{app:VMGdeta}, along with the method to obtain it.
In this framework, the matrix $T[\bm \theta]$ and the vector $V[\bm \theta]$ are thus analytically computed via automatic differentiation.
\jac{Details on the numerical regularization of the inversion of the matrix $T[\bm \theta]$ are presented in Appendix  \ref{app:regularization}}. 
\begin{figure*}[t!]
  \centering
    \includegraphics[width=\textwidth]{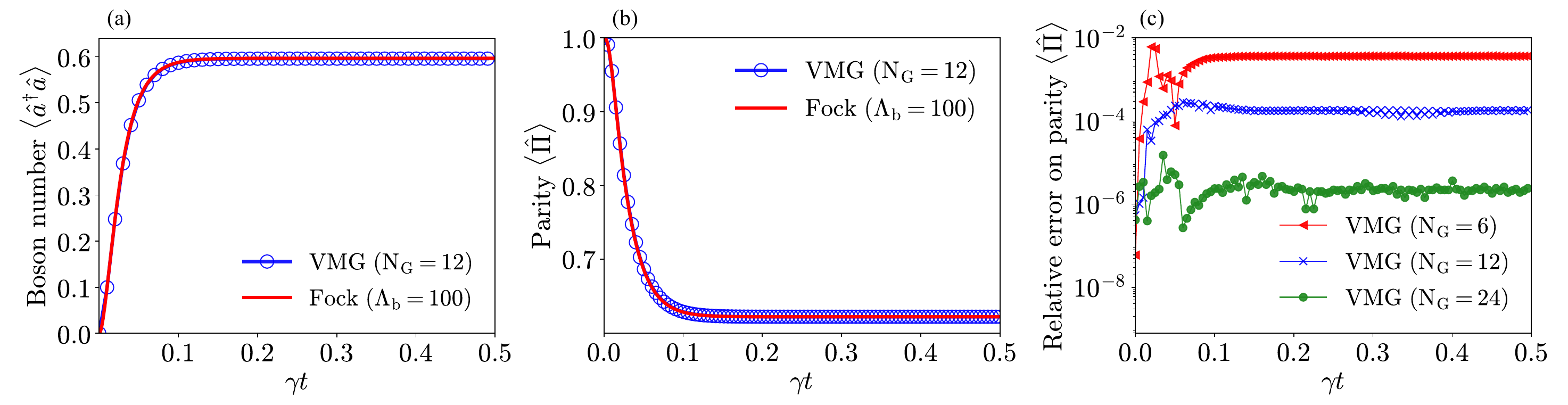}
  \caption{Dynamics of observables for the same open quantum system as in Fig.~\ref{cats}, but with a different set of illustrative parameters.
(a) Time evolution of the boson occupation number. The exact solution (solid red line) is compared to the VMG result (solid blue with marker)  obtained with 12 complex Gaussians.
(b) Same as in (a), but for the parity expectation value.
(c) Relative error (in log scale) of the dynamics of the parity expectation value. This relative error decreases $\textit{exponentially}$ with the  number of Gaussians used in the VMG ansatz.
Parameters: $ U/\gamma = 0.2, \  \Delta/\gamma = 0.1 $, $ \eta = 0 $, $ G/\gamma = 0.3 + 0.3i $. 
}
  \label{fig:one-mode-summary}
\end{figure*}

The ability to simulate the dynamics of a two-dimensional Bose–Hubbard system with two-boson coherent driving and two-body losses in the quantum critical regime for large lattice sizes highlights the efficiency of our phase-space method. The key breakthrough behind its accuracy and scalability lies in the full exploitation of the analytical properties of Gaussian functions through generalized Gaussian moments, combined with the use of state-of-the-art automatic differentiation and Taylor-mode expansion tailored to our physical setting. In contrast, previous approaches based on Gaussian ansätze in phase space \cite{Eeltink2023} were severely limited in both accuracy and scalability due to their reliance on purely numerical variational minimization, which neglected the analytical structure of Gaussian functions and did not leverage automatic differentiation—both of which are crucial for treating systems with a large number of modes.

Computing high-order moments via nested automatic differentiation poses significant memory challenges, which increases exponentially with the order of derivatives. 
As shown in Appendix \ref{appendix:phase_space}, for an Hamiltonian whose terms are at most of quartic order in the bosonic operators and for Lindblad jump operators that are at most quadratic in the bosonic operator, we have to deal at most with derivatives of order $4$. 
Moreover, the addition of the gradient with respect to the variational parameters $\bm \theta$ in Eq.~\eqref{eq:V_theta} increases, this maximal order of derivative, to the order $6$. 
To address this technical challenge, we have employed Taylor-mode automatic differentiation \cite{GriewankUtkeWalther2000}, a mathematically efficient—though still underutilized—technique \cite{dangel2025collapsingtaylormodeautomatic} based on expanding functions into their Taylor series (see Appendix~\ref{app:taylor-mode} for more details). This approach achieves at least factorial improvements in computational efficiency with respect to standard automatic differentiation, particularly with high-order derivatives.
By extracting the relevant Taylor coefficients, the required derivatives are obtained directly and simultaneously, making this method especially well-suited for the memory- and time-efficient computation of higher-order generalized Gaussian moments \cite{griewank2008evaluating}. In this work, we compute all such moments governing the dynamics in Eq.~\eqref{eq:V_theta} using Taylor-mode automatic differentiation as implemented in the \texttt{jax.experimental.jet} module of the JAX framework \cite{jax2018}.
Specifically, we employ Taylor mode automatic differentiation to compute the derivatives with respect to the vanishing currents $\bm J$ and $\tilde {\bm J}$ in Eq.~\eqref{eq:ggm}, while we adopt standard automatic differentiation with respect to the variational parameters $\bm \theta$ in the computation of the quantum geometric tensor $T$ and Liouvillian gradient $V$ in Eqs.~\eqref{eq:T_tensor} and~\eqref{eq:V_theta} respectively.

\section{Single Kerr Parametric Oscillator}\label{sec:single-kerr}

In order to benchmark our variational dynamical approach, we start by considering a bosonic system consisting of a single mode, where exact diagonalization can be performed. In particular, we study the Kerr parametric quantum oscillator, where bosonic Schrödinger cat states can emerge during the dynamics. This quantum model lies at the heart of several photonic quantum computing platforms \cite{Guillaud2019, Gravina2023, Hajr2024, Marquet2024, ding2025quantum}.
We explore the oscillator’s dynamics across a broad range of parameter regimes, including the deep quantum regime in which the Wigner function  displays negative fringes. 
In the reference frame rotating at half the pump frequency ($\hbar = 1$), the Hamiltonian of a single-site Kerr parametric oscillator reads
\begin{equation}
\hat H = \frac{U}{2} \hat{a}^{\dagger 2} \hat{a}^2+\frac{G}{2} \hat{a}^{\dagger 2}+\frac{G^*}{2} \hat{a}^2 - \Delta\hat a^\dagger \hat a .
\label{eq:single-mode}
\end{equation}
Here $\hat a$ is the bosonic annihilation operator ($[\hat a^\dagger, \hat a] = 1$), $U$ is the strength of the Kerr nonlinearity, and $G$ is the two-boson pump driving amplitude. The detuning is defined as $\Delta = \frac{\omega_{\rm p}}{2} - \omega_{\rm c}$, where $\omega_{\rm p}$ is the pump frequency and $\omega_{\rm c}$ the cavity frequency. We study the dynamics of the system under single- and two-boson losses, occurring at rates $\gamma$ and $\eta$, respectively. The corresponding jump operators are $\hat \Gamma_1 = \sqrt{\gamma} \hat a$ and $\hat \Gamma_2 = \sqrt{\eta} \hat a^2$.
For the single-mode case, we employed a brute-force exact diagonalization procedure using a Fock basis with a cutoff $\Lambda_{\rm b}$ on the maximum number of bosons. Simulations were performed using the open-source software QuTip \cite{qutip}, and convergence was carefully checked by increasing the cutoff.
\begin{figure}
    \centering
    \includegraphics[width=\linewidth]{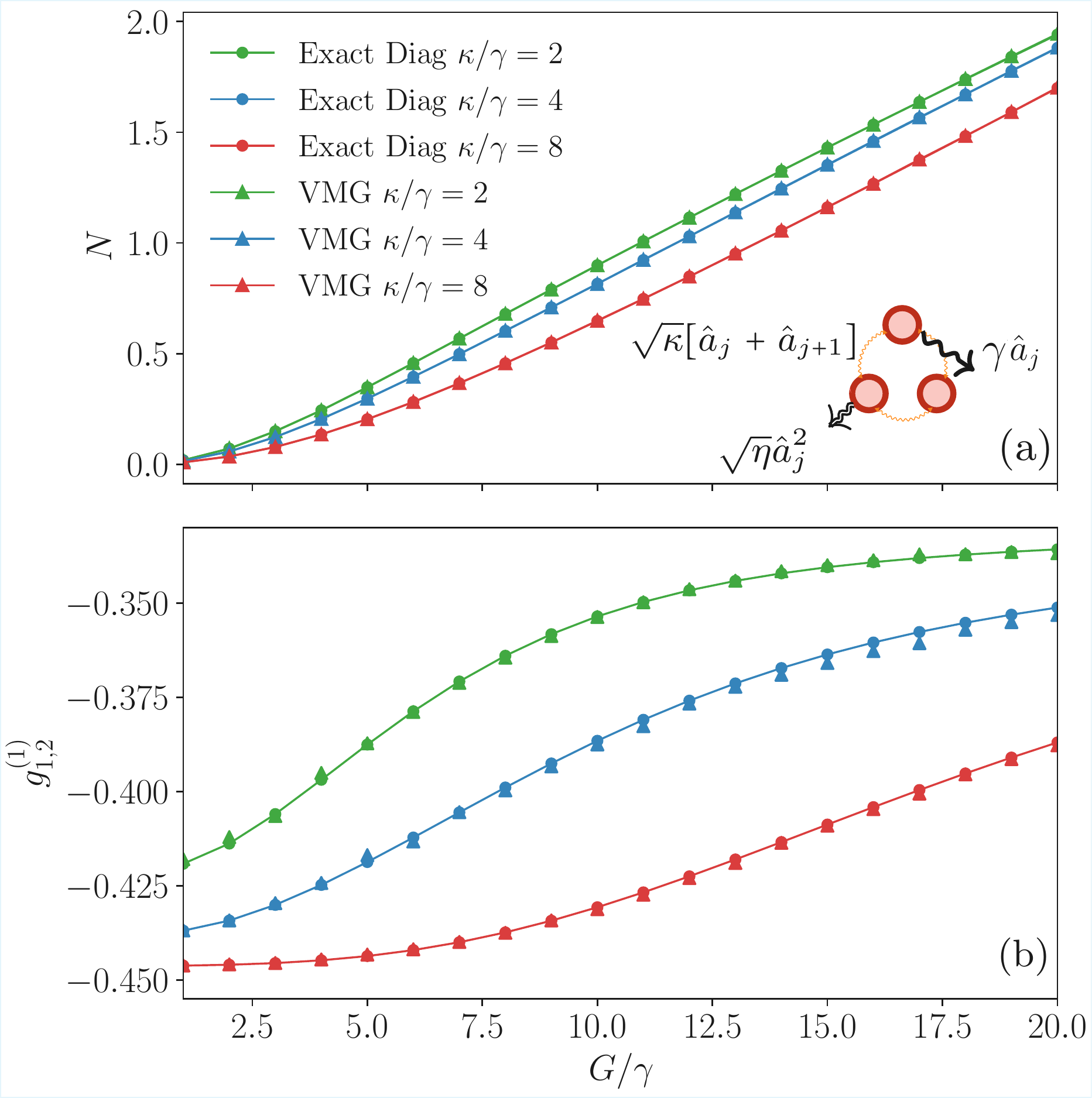}
    \caption{\cri{(a) Steady-state boson number $N$ and (b) two-point correlation function $g^{(1)}_{1,2}$ for a system composed of three quantum parametric Kerr oscillators coupled dissipatively via non-local quantum jump operators (see inset of panel (a)). The solid lines with circular markers represent results obtained via exact diagonalization, while the triangular markers show the results from the variational phase-space approach ($N_G = 128$). Different colors correspond to different values of the non-local loss rate $\kappa$. Exact diagonalization was performed using a Fock state basis with a cutoff of $10$ boson per site. Parameters in the simulations were $\eta/\gamma=1$ and $U/\gamma=10$}.}
\label{fig:n_photon_g12}
\end{figure}

\begin{figure*}[t!]
  \centering
  \includegraphics[width=0.8\textwidth]{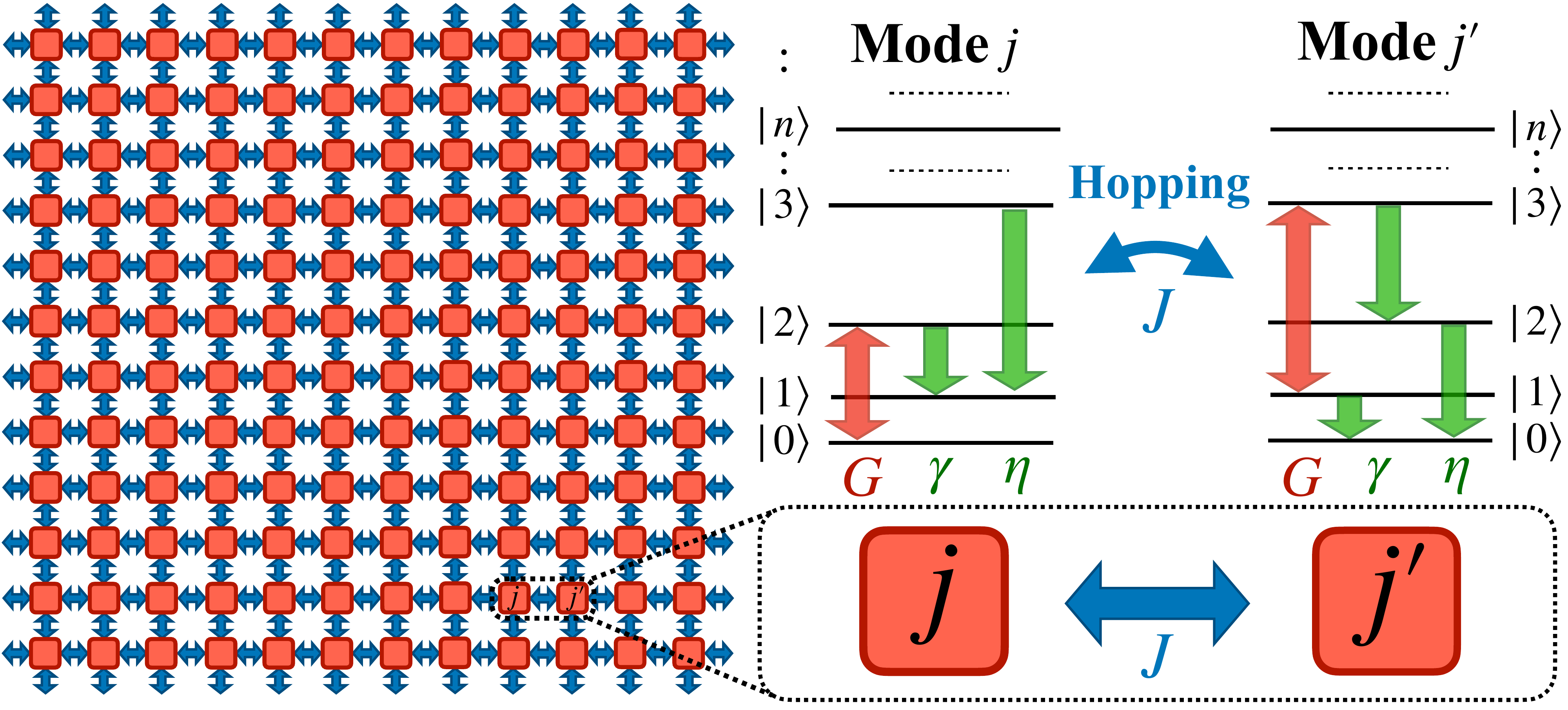}
  \caption{Scheme of the considered $12\times 12$ Bose-Hubbard lattice system with periodic boundary conditions with two-boson driving and losses. The term $G$ stands for the two-boson driving amplitude, while $\gamma$ and $\eta$ represent the single- and two-boson loss rates respectively. Nearest-neighbor sites (modes) are coupled by the boson hopping quantified by $J$. The energy levels of the Fock states of each mode are not equally spaced reflecting the presence of the on-site boson-boson interaction $U$.}
  \label{fig:scheme_2D_lattice}
\end{figure*}

In the steady state, the Wigner function for a Kerr parametric oscillator is always positive \cite{Bartolo2016}. Nevertheless, the Wigner function, in a regime with a small single-boson loss rate $\gamma/\eta \ll 1$, can display regions with deep Wigner negativities during the temporal transient.
As shown in Fig.~\ref{cats}(a), realized for $\gamma = 0$, the time evolution of the Wigner function, reconstructed via exact diagonalization (Fock cutoff $\Lambda_{\rm{b}} = 100$), displays negative values. This snapshot of the Wigner function is realized at time $\eta t=5$, i.e. well before the system has relaxed to the steady state.
As it can be seen in Fig.~\ref{cats}(b)-(c), a VMG ansatz with $32$ complex Gaussian functions excellently reproduces the dynamics of the Wigner function in a regime where it is highly non-Gaussian and with deep negativities. Panels (a) and (b) are not distinguishable by eye. Panel (c) displays the negligible difference on a color-scale reduced by several orders of magnitude. With the Wigner function, \fra{the expectation value $\langle \hat O\rangle = \rm tr[\hat O \hat \rho]$ of any observable $\hat O$ can be computed}. As detailed in Appendix \ref{appendix:phase_space}, this is achieved analytically by integrating the corresponding Weyl symbols $O_{\rm{W}}$ of the observable over the Multi-Gaussian Wigner function $W_{\bm \theta}(\bm \xi )$ for any time $t$. \\
Specifically, in Fig.~\ref{fig:one-mode-summary}(a)-(b) we report the time evolution of the expectation values for the boson number
$\hat N \equiv \hat a^\dagger \hat a$ and the parity operators $\hat \Pi \equiv \exp(i \pi \hat a^\dagger \hat a)$.
For this given Hamiltonian, already for a small number of Gaussians in the ansatz, such as $N_{\rm G}=6$, the agreement with the exact dynamics is outstanding. Remarkably, the relative error on the observables can be exponentially reduced by increasing the number of Gaussian functions $N_{\rm G}$ in Eq.~\eqref{eq:vmg}, as shown with a log scale in Fig.~\ref{fig:one-mode-summary}(c). In particular, by increasing the number of Gaussians by a factor $4$, the relative error decreases by approximately $4$ orders of magnitudes. 

\fra{\section{Parametric oscillators with non-local losses}
\label{sec:Kerr-nonlocal}

To further benchmark our variational phase-space method, we apply it to a system of bosonic modes coupled via non-local quantum jump operators. Beyond standard local losses, photonic platforms can realize such engineered non-local dissipation, which induces effective inter-mode interactions and rich many-body dynamics. We compare our variational results against exact diagonalization for the model studied in Ref.~\cite{Li2021Dissipation}, governed by the effective Hamiltonian:

\begin{equation}
\hat{H}_{\mathrm{eff}} =
\sum_{j} 
    \frac{U}{2}\hat{a}_{j}^{\dagger 2}\hat{a}_{j}^{2}
    + \frac{G}{2}\hat{a}_{j}^{\dagger 2}
    + \frac{G^{*}}{2}\hat{a}_{j}^{2}.
\end{equation} 
Analogously to Eq.~\eqref{eq:single-mode}, $U$ represents the Kerr nonlinearity strength and $G$ the two-boson pumping amplitude. In addition to the local losses present in the single-mode oscillator ($\hat{\Gamma}_{1,j} = \sqrt{\gamma} \hat{a}_j$ and $\hat{\Gamma}_{2,j} = \sqrt{\eta} \hat{a}^2_j$), the system features non-local jump operators of the form $\hat{\Gamma}_{\kappa,j} = \sqrt{\kappa} (\hat{a}_j+\hat{a}_{j+1})$, which generate collective correlations.

As shown in Fig.~\ref{fig:n_photon_g12}(a), our method accurately reproduces single-body observables, such as the average boson population
\[
N = \frac{1}{L}\sum_{j=1}^L \langle \hat{a}_j^\dagger \hat{a}_j \rangle,
\]
for a chain of $L=3$ coupled cavities. Furthermore, Fig.~\ref{fig:n_photon_g12}(b) displays the nearest-neighbor correlation function:
\begin{equation}
    g^{(1)}_{1,2} = \frac{1}{LN}\sum_{j=1}^L \langle \hat{a}_j^\dagger \hat{a}_{j+1} \rangle .
\end{equation}
As reported in Fig. \ref{fig:n_photon_g12}, we observe that the correlations approach $-1/3$, which coincides with the spin-correlation value of the antiferromagnetic triangular Ising model. Capturing these effective spin frustration effects requires a slightly larger variational basis $N_{\mathrm{G}}$ than in the single-mode case discussed in Sec.~\ref{sec:single-kerr}; specifically, we employed $N_{\mathrm{G}}=128$ Gaussians (specifically $64$ couples) for each value of non-local loss $\kappa$. We detail the parametrization and initialization of the Gaussian ansatz in Appendix~\ref{app:parametrization}.
}
\section{Bose-Hubbard lattices with two-boson driving and dissipation}
\label{sec:bose-hubbard-steady}

In this Section, we will consider a nontrivial system: the Bose-Hubbard model with periodic boundary conditions subject to two-boson coherent driving, as well as both single-boson and two-boson losses. This system, pictured in Fig.~\ref{fig:scheme_2D_lattice}, can be viewed as a lattice of Kerr parametric quantum oscillators, with each site coupled to its nearest neighbors via hopping. In the previous section, we have considered the system consisting of a single site.\\
Networks of coupled Kerr parametric oscillators have garnered significant attention due to their ability to form interacting lattices, offering a promising platform for quantum information processing \cite{aspuru-guzik2012photonic}. In particular, these systems—alongside optical parametric oscillators \cite{Tosca2025}—have been proposed as hardware-efficient solvers for the Ising problem and as intrinsically noise-resilient architectures for quantum computation \cite{Leghtas2015}.
The Hamiltonian of the corresponding model reads  ($\hbar = 1$)
\begin{equation}
\begin{aligned}
\hat{H} = & \sum_{j=1}^{M} \left[-\Delta\, \hat{a}_j^\dagger \hat{a}_j + \frac{U}{2} \hat{a}_j^{\dagger 2} \hat{a}_j^2 + \frac{G}{2} \hat{a}_j^{\dagger 2} + \frac{G^*}{2} \hat{a}_j^2 \right] \\
& - \sum_{\langle j, j' \rangle} \frac{J}{z} \left(\hat{a}_j^\dagger \hat{a}_{j'} + \hat{a}_{j'}^\dagger \hat{a}_j \right),
\end{aligned}\label{eq:driven-diss-Bose-Hubb}
\end{equation}
\begin{figure*}[t!]
  \centering
    \includegraphics[width= 0.9\textwidth]{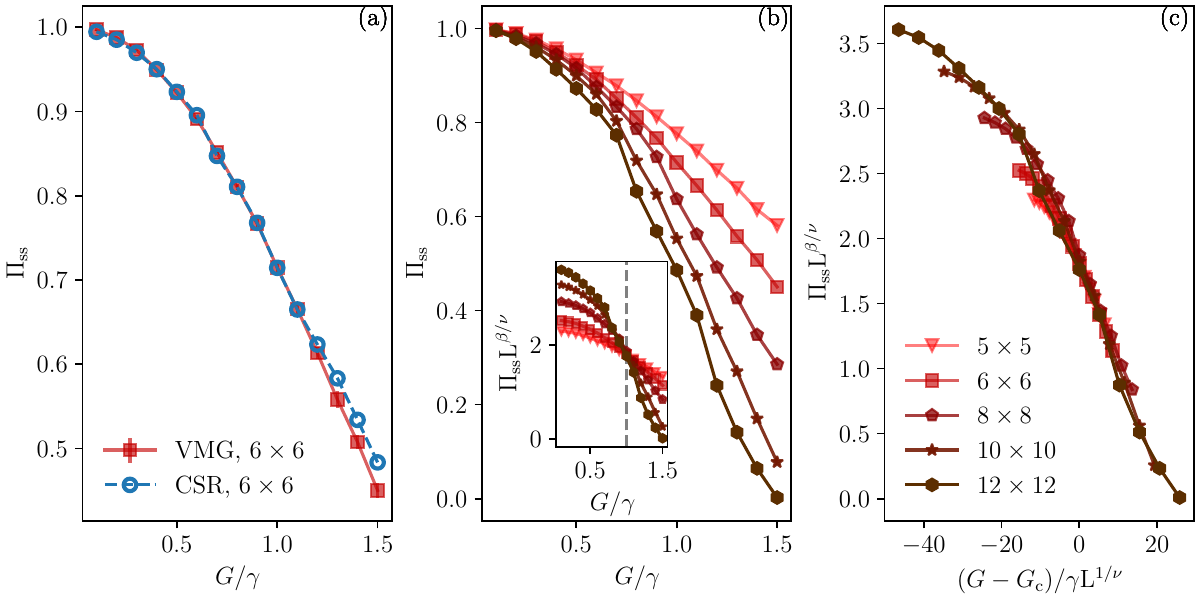}
    \caption{(a) Comparison between the steady-state expectation value of the global parity operator obtained using both the Corner Space Renormalization method (CSR) \cite{rota2019quantum} and our variational approach with a Variational Multi-Gaussian (VMG) ansatz, employing $N_{\rm G} = 16$ Gaussians for a 2D driven-dissipative Bose-Hubbard lattice Eq.~\eqref{eq:driven-diss-Bose-Hubb}.  
(b) Steady-state parity expectation values computed with our variational approach ($N_{\rm G} = 16$) for increasing values of $L$, where $L \times L$ is the number of lattice sites. The parity is plotted as a function of the driving amplitude $G/\gamma$. In the inset, the critical pump amplitude $G_{\rm c}$ (dashed vertical line) is estimated to be at $G_{\rm c} \simeq 1.0 \gamma$.  
(c) Rescaled steady-state expectation values of the parity operator as a function of the rescaled pump amplitude.  
The critical exponents $\beta$ and $\nu$ employed are those of the 2D quantum Ising model in a transverse field: $\beta = 0.32641871$, $\nu = 0.62997097$ \cite{Chang2025Bootstrapping,Reehorst2022Bounds,Komargodski2017RandomBond}.
The critical value of the two-boson pump amplitude is $G_{\rm c} \approx 1.0 \gamma$.  
Parameters:  
$\Delta/\gamma = -20.0$, $\eta/\gamma = 1.0$, $U/\gamma = 40.0$, and $J/\gamma = 20.0$.
 \label{fig:parity_scaling}}
\end{figure*}
where $\hat{a}_j$ denotes the bosonic annihilation operator of the $j-$th mode (site), $U$ is the Kerr nonlinearity, $\Delta$ the detuning, $G$ the two-boson drive amplitude, and $J$ the nearest-neighbor hopping rate normalized by the coordination number $z$. Single- and two-boson losses are incorporated via the dissipation superoperator $\mathcal{D}$ acting on the density matrix $\hat{\rho}(t)$:

\begin{equation}\label{eq:dissipation}
    \mathcal{D}[\hat{\rho}(t)] = \sum_{k=1}^{2} \sum_{j=1}^{M} \left[
    \hat{\Gamma}_{k,j} \hat{\rho}(t) \hat{\Gamma}_{k,j}^\dagger - \frac{1}{2} \left\{ \hat{\Gamma}_{k,j}^\dagger \hat{\Gamma}_{k,j}, \hat{\rho}(t) \right\} \right],
\end{equation}
where $M$ denotes the number of lattice sites, and the jump operators on site $j$ are given by 
\begin{equation} \hat{\Gamma}_{1,j} = \sqrt{\gamma} \, \hat{a}_j \hspace{0.5cm}\mbox{and}\hspace{0.5cm}
\hat{\Gamma}_{2,j} = \sqrt{\eta} \, \hat{a}_j^2 , \end{equation} corresponding respectively to single- and two-boson loss processes.

To the best of our knowledge, a finite-size scaling analysis of this model has been studied only in the steady state \cite{rota2019quantum}, using the Corner Space Renormalization method \cite{Finazzi2015Corner}, which is a technique to find the stationary states of low-entropy open quantum systems where the basis of states is judiciously constructed through a spatial renormalization procedure. The Corner Space Renormalization has controllable accuracy, as in exact diagonalization approaches, but the lattice size that can be simulated is limited by the memory of the calculated density matrix, which depends on its entropy. \cri{Indeed, in corner-space renormalization one must store in memory not only the density matrix of size $N_{\mathrm{states}} \times N_{\mathrm{states}}$, but also the Hamiltonian matrix and the jump operators, one for each lattice site.}

Here, we will benchmark our predictions for the steady-state observables against the results obtained with the Corner Space Renormalization method \cite{rota2019quantum}. We will extend the analysis to much larger lattice sizes \cri{and most importantly we will be able to determine the time-dependent dynamics}. 

Note that in this work we determine the steady state by evolving the master equation up to sufficiently long times. Alternatively, it could be obtained variationally by directly solving the equation $\mathcal L \rho_{\rm ss} = 0$ \footnote{This approach would require constrained-optimization techniques to ensure that the state remains physical, i.e., positive semidefinite and with unit trace. In contrast, for time evolution, if the ansatz is sufficiently expressive (i.e., includes a large enough number of Gaussians) and the initial density matrix is physical, the Lindblad master-equation dynamics preserves the positive-semidefinite character of the density matrix.}. 

The Liouvillian of the considered system, with Hamiltonian in Eq.~\eqref{eq:driven-diss-Bose-Hubb} and dissipator in Eq.~\eqref{eq:dissipation}, is invariant under the transformation $\hat{a}_j \xrightarrow {} - \hat{a}_j $ and $\hat{a}^{\dagger}_j \xrightarrow {} - \hat{a}^{\dagger}_j $. 
This means that the even parity of a Wigner function in phase space $W(\boldsymbol{q},\boldsymbol{p}) = W(-\boldsymbol{q},-\boldsymbol{p})$ is preserved by the time evolution described by the master equation. 
Such non-equilibrium Bose-Hubbard model undergoes a dissipative quantum phase transition \cite{rota2019quantum} for a critical value of the two-boson driving amplitude $G$,  associated with a spontaneous symmetry breaking of such discrete  $\mathbb Z_2$ symmetry of the Liouvillian in the thermodynamical limit. 

A relevant quantity to characterize this phase transition is the boson number parity operator defined as:
\begin{equation}
\hat \Pi= \exp \left (i\pi \sum_{j=1}^M \hat a_j^\dagger \hat a_j  \right ).
\end{equation}
The emergence of the above mentioned phase transition with increasing lattice size can be witnessed by the finite-size scaling of the expectation value of this parity operator in the steady state \cite{rota2019quantum}. 
As detailed in Appendix \ref{appendix:phase_space}  the expectation value of the parity operator is proportional to the value of the Wigner function at the origin of the phase space. 

Indeed, as illustrated in Fig.~\ref{fig:parity_scaling}, our variational method faithfully reproduces the steady‑state parity across the second‑order phase transition. In particular, we have achieved an excellent agreement between our results and those obtained via Corner Space Renormalization. As an illustrative example of the successful benchmarking, we report the results for a $6\times6$ lattice in Fig.~\ref{fig:parity_scaling}(a), where we plot the steady-state expectation value of the parity operator versus the driving $G$. \cri{In Ref.~\cite{rota2019quantum}, the largest steady-state calculations performed with the corner-space renormalization method required a Hilbert-space dimension of about $N_{\mathrm{states}} \sim 5000$. The corresponding total memory cost, due to the corner-space algorithm, was on the order of $10^9$ complex numbers ($\approx N_{sites}^2   \times N_{\mathrm{states}}$). In contrast, the present approach allows us to simulate not only the full time dynamics but also significantly larger lattices while requiring far less memory. In the largest dynamical simulations reported here, corresponding to $12 \times 12$ lattices, the maximum number of real variational parameters is approximately $7000$. Further details about the complexity and the scaling of the variational parameters is reported in Appendix \ref{app:parametrization}.}

In Fig.~\ref{fig:parity_scaling}(b), we report the parity expectation value versus the two-boson driving $G$ for increasing lattice sizes. Our finite size-scaling analysis shows an emergent phase transition belonging to the universality class of the 2D quantum Ising model. Indeed, as shown in Fig.~\ref{fig:parity_scaling}(c), a rescaling \cite{fisher1989boson,cardy1996scaling} of the parity expectation value and of the driving $G$ with the system size $L$ and the critical exponents $\beta$ and $\nu$, all the curves nicely collapse around the critical driving amplitude $G_{\rm c}$. As shown in the inset of Fig.~\ref{fig:parity_scaling}(b), the value of the critical driving amplitude $G_{\rm c}$ can be extracted by the crossing point obtained by rescaling only the parity.

As the system size increases, the  time needed to reach the steady state correspondingly grows. For the results reported in this work, we evolved the Lindblad master equation up to a final time of $20\gamma^{-1}$, $40\gamma^{-1}$, $60\gamma^{-1}$, $120\gamma^{-1}$, and $120\gamma^{-1}$ for lattice sizes $5\times5$, $6\times6$, $8\times8$, $10\times10$, and $12\times12$, respectively. Note that the calculations for lattices up to $10\times10$ were carried out on a single Nvidia A100 GPU, while those for the $12\times12$ lattice were performed on a single Nvidia H100 GPU. 
\section{Critical slowing down and dynamical critical exponents}\label{sec:bose-hubbard-dynamics}

\begin{figure}[t!]
    \centering
    \includegraphics[width=\linewidth]{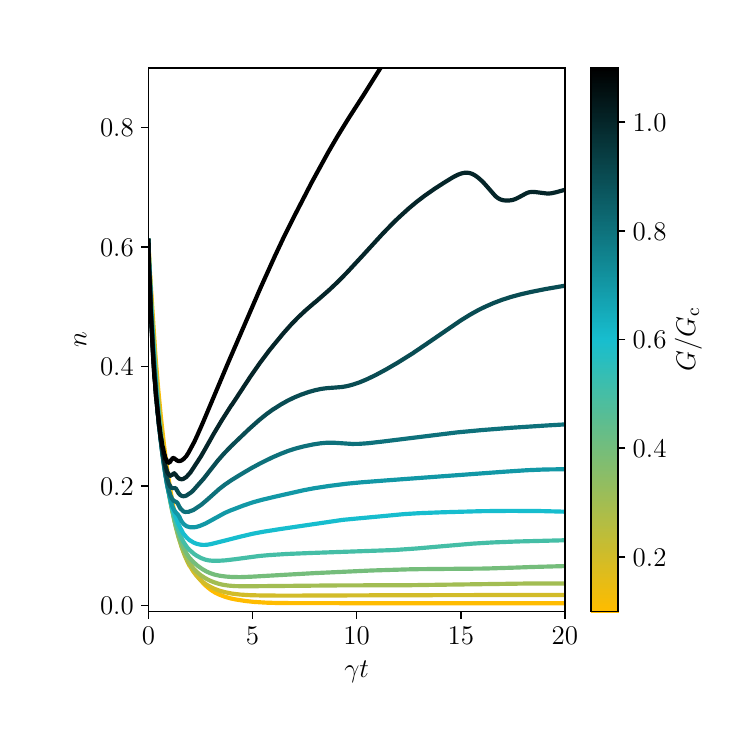}
    \caption{Time evolution of the average boson number $n$ for a $12 \times 12$ Bose-Hubbard lattice. As the drive amplitude $G$ increases, the relaxation dynamics becomes significantly slower. \fra{Same parameters of the Liouvillian $\mathcal L$ as in Fig.~\ref{fig:parity_scaling}. Here the initial state is chosen away from the bosonic vacuum, as described in detail in Appendix \ref{app:parametrization}, where this choice for the initial state is deemed as $W_{\rm init,1}$.}
    }
    \label{fig:12x12dyn}
\end{figure}
\begin{figure*}
  \centering
    \includegraphics[width=1\textwidth]{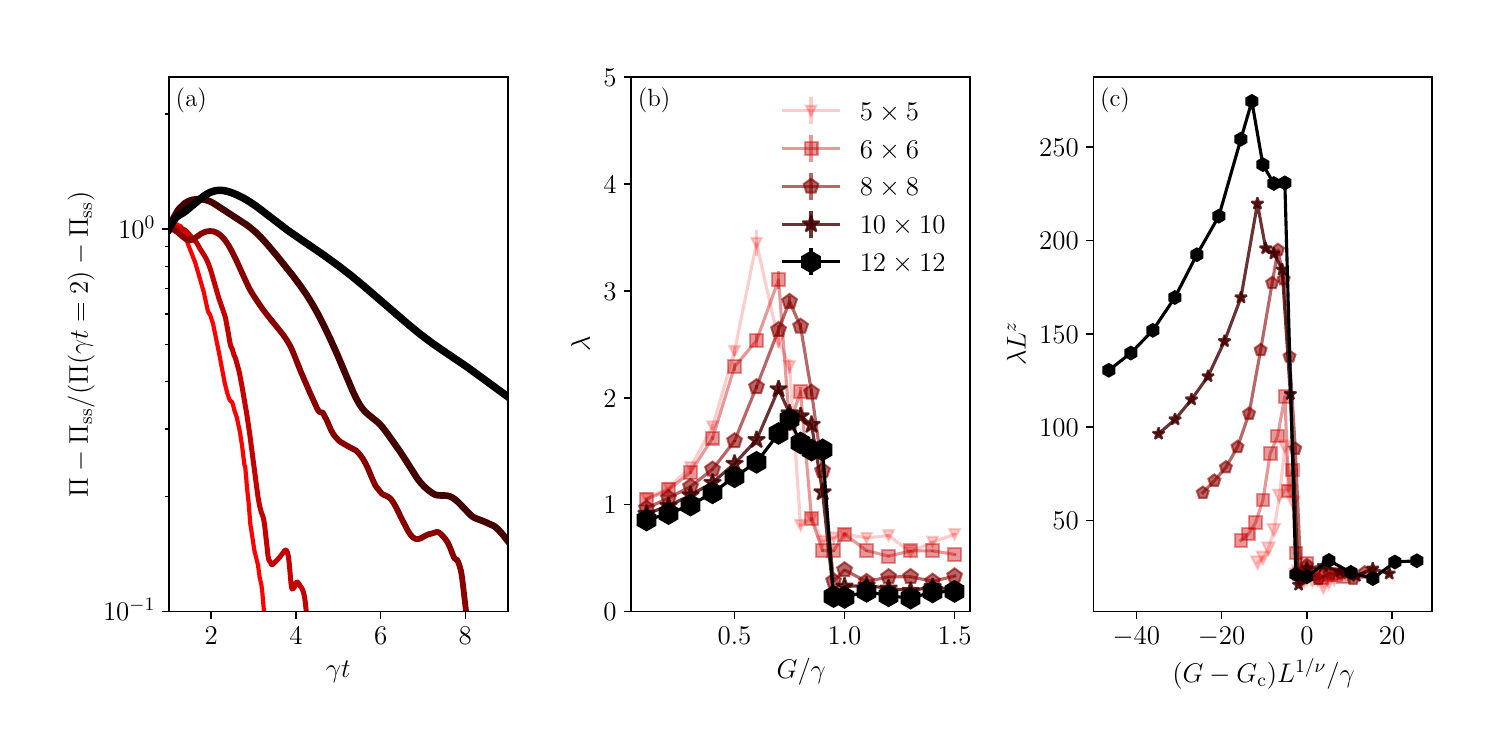}
    \caption{
    (a) Illustrative example of the temporal dynamics of the parity operator expectation value, shown on a log scale, at the critical driving amplitude $G = G_{\rm c} \sim 1.0 \ \gamma$ for different lattice sizes (see legend in panel b).
    (b) Asymptotic decay rate $\lambda$ toward the steady state as a function of the driving amplitude $G$, shown for different lattice sizes $L \times L$.  
    (c) Finite-size scaling analysis revealing dynamical critical behavior consistent with the 2D quantum Ising universality class, with critical exponent $z = 2.0235$. By rescaling the shifted two-boson driving amplitude $(G - G_{\rm c})/\gamma$ with the rescaled lattice side length $L^{1/\nu}$, the rescaled decay rates $\lambda L^z$ collapse onto a single universal curve for all considered system sizes. Same parameters as in Fig.~\ref{fig:parity_scaling}.
    \label{fig:gap}}
\end{figure*} 
In this section, we apply our variational method to calculate the dynamical properties of the second-order quantum phase transition in the 2D network of Kerr parametric quantum oscillators introduced in the previous section. The study of critical dynamical quantities is challenging because it requires the ability to obtain highly accurate results on large-scale lattices over diverging time scales.  In order to characterize the critical dynamics, we focus here on the asymptotic decay rate. 
This is the rate at which observables approach their steady-state value in the long-time limit. At the critical point, the asymptotic decay rate is expected to vanish, corresponding to a critical slowing down of relaxation \cite{Minganti2018Spectral}. The asymptotic decay rate is also called the Liouvillian frequency gap, as it can be computed from the spectrum of the Liouvillian superoperator. 

To the best of our knowledge, this is the first time the dynamics of such a non-equilibrium phase transition is explored in the quantum regime.  Previous studies in the literature have investigated first-order phase transitions in the semiclassical regime of the Bose-Hubbard model with single-boson driving \cite{Vicentini2018}, where the Wigner function remains positive and can be simulated using the truncated Wigner approximation. Another work \cite{Rota2020} considered single-Gaussian trajectories to study the driven-dissipative 2D Bose-Hubbard model with two-boson driving, but in a semiclassical regime dominated by single-boson losses (indeed, the extracted critical exponents correspond to the universality class of the classical 2D Ising model).

This critical slowing down of the dynamics for the considered model can be appreciated in Fig.~\ref{fig:12x12dyn}, where, for a $12\times 12$ lattice, we show the dynamical evolution of the expectation value of the mean boson number $n = \langle\hat n\rangle$:
\begin{equation}
    \hat n= \frac{1}{L^2}
    \sum_{j=1}^{L^2}
    \hat a_j^\dagger \hat a_j,
\end{equation}
where $L^2$ is the number of sites of the square $L \times L$- lattice.
We refer to Appendix \ref{app:parametrization} for a detailed description of the initial conditions used for the calculated dynamics.

By examining Fig.~\ref{fig:12x12dyn}, one observes that as the two-boson driving amplitude $G$ approaches the critical value $G_{\rm c}$, the time required to reach the steady state becomes progressively longer. Specifically, while for $G/G_{\rm c} \ll 1$ the system relaxes well before time $t=20\gamma^{-1}$, for $G=G_{\rm c}$ the dynamics is still far from the steady state after the same time. This diverging time scale highlights how the variational approach captures the vanishing of the gap $\lambda$ near the critical point $G_{\rm c}$. As already mentioned, the relaxation time near the critical driving further increases when increasing the system size, as shown in Fig.~\ref{fig:gap}(a).

In Fig.~\ref{fig:gap}(b), we report the behavior of the Liouvillian gap for different system sizes ranging from $25$ to $144$ bosonic modes.  
As expected for second-order phase transitions \cite{Minganti2018Spectral}, the gap tends toward vanishingly small values near the critical point, leading to a diverging relaxation time in the thermodynamic limit.  

Remarkably, in Fig.~\ref{fig:gap}(c), we demonstrate that the driven-dissipative Bose-Hubbard model belongs to the Ising universality class by rescaling with the corresponding dynamical critical exponent $z = 2.0235$ \cite{Adzhemyan2022dynamic}. Indeed, when we plot the rescaled asymptotic decay rates $\lambda L^z$ as a function of $(G-G_{\rm c}) L^{1/\nu}/\gamma$, the data collapse onto a single universal curve in the critical region, which features a sharp drop of the decay rates, which for any value  $G > G_c$  vanish in the thermodynamic limit.

\section{Conclusions and Outlook}
\label{conclusions}
We have introduced a scalable variational method for simulating the dynamics of interacting open quantum bosonic systems deep in the quantum regime. Our approach is based on a multi-dimensional Wigner phase-space representation and employs a Variational Multi-Gaussian (VMG) ansatz, whose accuracy is systematically controlled by the number of Gaussian components. The equations of motion are derived from the Dirac–Frenkel principle and are evaluated efficiently by leveraging the analytical structure of Gaussian functions together with automatic differentiation.

This framework enables the simulation of dissipative many-body quantum dynamics in strongly correlated regimes, where standard approaches become intractable due to large Hilbert space sizes or strong Wigner negativities. As a key application, we investigated the driven-dissipative two-dimensional Bose–Hubbard model with two-boson coherent driving and two-body losses, accessing its quantum critical dynamics far from equilibrium. By analyzing the time-dependent relaxation and performing finite-size scaling of the Liouvillian spectral gap, we revealed a vanishing asymptotic decay rate and extracted critical exponents belonging to the 2D quantum Ising universality class.

Beyond fundamental insights into strongly interacting bosonic systems, our method provides a powerful computational tool for designing and simulating complex networks of strongly interacting bosonic modes—such as nonlinear quantum oscillators—used in quantum information processing. The ability to model such architectures in the presence of dissipation and strong quantum correlations opens new avenues for the development of robust quantum technologies. Looking forward, our framework based on automatic differentiation may be extended to other analytical ansätze and to phase-space representations tailored to spins and fermions, enabling broad applications across condensed matter, quantum optics, and quantum engineering. \\

\acknowledgements 
We acknowledge support from the French ANR project FracTrans (grant ANR-24-CE30-6983) and  a grant (Polaritonic) from the French Government managed by the
ANR under the France 2030 programme with the reference ANR-24-RRII-0001.
This work was granted access to the HPC resources of IDRIS under the allocation 2025-AD010612462R2 made by GENCI.
Numerical computations were partly performed on the S-CAPAD/DANTE platform, IPGP, France.
We also thank Zejian Li for providing data for the frustrated coupled photonic resonators.

\begin{widetext}
\appendix
\section{Generalized Gaussian moments}
\label{app:VMGdeta}
As mentioned in the main text, the differential Liouvillian operator $\mathcal{L}_W$ can be decomposed as a sum of multiple terms $\mathcal L_{W,\mathbf{I}}$, such that $\mathcal{L}=\sum_\mathbf{I} \mathcal L_{W,\mathbf{I}}$, all of the form:
\begin{equation}
    \mathcal L_{W,\mathbf{I}} \propto 
    \xi_{i_1}...\xi_{i_l}
    \partial_{\xi_{j_1}}...\partial_{\xi_{j_k}}.
\end{equation}
To unravel the time evolution of the variational parameters $\bm \theta$ that appear in the variational Wigner function $W_{\bm \theta}(\bm \xi)$, one needs to compute the Liouvillian gradient vector $\bm V$ in the equation Eq.~\eqref{eq:DiracFrank}. 
When the Wigner function is given by the sum of multiple Gaussian functions, this can be done analytically via automatic differentiation. 
To simplify the notation, we consider the case of Gaussian functions in which the normalization coefficient $c=1$ (cf. Eq. \eqref{eq:Gaussian}). In this case a Gaussian function is identical to the multivariate normal distribution $\mathcal N$ with centers $\bm \mu$ and covariance matrix $\Sigma$: $G(\bm \xi;\bm \theta)|_{c=1} \equiv \mathcal N (\bm \xi;\bm \mu, \Sigma)$.
The normal distribution is given as:
\begin{equation}
\begin{aligned}
    &\mathcal N (\boldsymbol \xi ; \boldsymbol{\mu}, \Sigma)
    = 
    \frac{1}{\sqrt{(2\pi)^{M} \det(\Sigma)}} \exp\left[{-\frac{1}{2} (\boldsymbol\xi-\boldsymbol{\mu}) \Sigma^{-1}(\boldsymbol\xi-\boldsymbol{\mu}) } \right].
\end{aligned}
\end{equation}
One can start by writing the integral of the terms $\mathcal L_{\mathbf I}$, on a set of normalized Gaussians, in the following fashion:
\begin{equation}
\begin{aligned}
    &\llangle
    \xi_{i_1}...\xi_{i_l}
    \partial_{\xi_{j_1}}...\partial_{\xi_{j_k}}
    \rrangle_{(\bm \mu,\Sigma ),( \bm \mu',\Sigma')}
    \equiv \\
    &\int d^{2M} \bm \xi \ 
    \N(\bm \mu,\Sigma)
    \xi_{i_1}...\xi_{i_l}
    \partial_{\xi_{j_1}}...\partial_{\xi_{j_k}}
    \N( \bm \mu',\Sigma').
\end{aligned}
\end{equation}
By inserting in the previous equation a source term $e^{\bm J^T \bm \xi}$ and deriving with respect to $\bm J$,
the quantity $\xi_{i_1}...\xi_{i_k}$ 
can be factored out of the integral as:
\begin{equation}
\begin{aligned}
    &\int d^{2M} \bm \xi \ 
    \N( \bm \mu,\Sigma )\,
    \xi_{i_1} \dots \xi_{i_l} \,
    \partial_{\xi_{j_1}} \dots \partial_{\xi_{j_k}}
    \N( \bm \mu',\Sigma' )
    = \\[0.5em]
    &\quad \partial_{J_{i_1}} \dots \partial_{J_{i_l}}
    \int d^{2M} \bm \xi \ 
    e^{\bm J^T \bm \xi} \,
    \N( \bm \mu,\Sigma ) \times \\[0.3em]
    &\quad\quad
    \partial_{\xi_{j_1}} \dots \partial_{\xi_{j_k}}
    \N( \bm \mu',\Sigma' ) 
    \Big|_{\bm J = 0}.
    \label{eq:partial_xi}
\end{aligned}
\end{equation}
Now, let us notice that the term $e^{\bm J^T \bm \xi}\N(\bm \mu,\Sigma)$, appearing on the left hand side in the previous equation, can be recast as:
\begin{equation}\label{eq:Gauss_decomp}
    e^{\bm J^T \bm \xi}\N(\bm \mu,\Sigma)
    =
    e^{\frac{1}{2}\bm J^T \Sigma \bm J+\bm J^T \bm \mu} \N(\bm \mu + \bm J^T \Sigma,\Sigma).
\end{equation}
Not being dependent on any phase-space variable $\bm \xi$, the factor $e^{\frac{1}{2}\bm J^T\Sigma \bm J+\bm J^T \bm \mu}$ can be safely taken out of the integral in Eq.~\eqref{eq:partial_xi}. 
Additionally, the term $\partial_{\xi_{j_1}}...\partial_{\xi_{j_k}}$, on the right-hand side of Eq. \eqref{eq:partial_xi}, can be treated analogously to ${\xi_{i_1}}...{\xi_{i_l}}$ by rewriting the Gaussian function in the Fourier dual space.
Specifically the Fourier dual of $\mathcal{N}(\bm \mu, \Sigma)$ is defined as:
\begin{equation}
    \N(\bm \mu,\Sigma)
    =
    \int d \bm \kappa
    e^{2\pi i \bm \xi^T \bm \kappa}
    \tilde{\mathcal N}(\bm \mu,\Sigma)[\bm \kappa],
\label{eq:fourier_gaus}
\end{equation}
where $\tilde \N(\bm \mu,\Sigma)[\bm \kappa]$ explicitly reads as:
\begin{equation}
    \tilde \N(\bm \mu,\Sigma)[\bm \kappa]   
    =
    e^{-2\pi i \bm \mu^T \bm \kappa}
    e^{-2\pi^2 \bm \kappa^T \Sigma \bm \kappa}.
    \label{eq:inv_fourier_gaus}
\end{equation}

Under such transformation, and introducing for simplicity the notation $\bm \nu = \bm \mu + \bm J^T \Sigma$, the integrand on the right-hand side of Eq. \eqref{eq:partial_xi} becomes: 
\begin{equation}
\begin{aligned}
   &\int d^{2M} \bm \xi \
    \N(\bm  \nu,\Sigma)
    \partial_{\xi_{j_1}}...\partial_{\xi_{j_k}}
    \N( \bm \mu',\Sigma')
    =\\
    &\quad
    \int d \bm \xi d \bm \kappa ' d \bm \kappa e^{2\pi i \bm \xi^T (\bm \kappa+\bm \kappa')}
    \tilde G(\bm \nu,\Sigma )[\bm \kappa'] \times \\
    &\qquad (2\pi i)^n\kappa_{{j_1}}...\kappa_{ \xi_{j_k}}
    \tilde G(\bm \mu',\Sigma')[\bm \kappa]=\\
    & 
    \partial_{\tilde J_{j_1}}...\partial_{\tilde J_{j_n}}
    \int d \kappa
    \tilde \N( \bm\nu,\Sigma )[-\bm \kappa]
    e^{2\pi i \tilde {\bm J}^T \bm \kappa}
    \tilde \N( \bm \mu',\Sigma')[\bm \kappa],
\end{aligned}
\label{eq:fourier_int}
\end{equation}
where we have introduced a (dual) source term $e^{2\pi i \tilde {\bm J}^T\bm \kappa}$, in the same flavor of what already done in direct space. The integral term in Eq.~\eqref{eq:fourier_int} can then be rewritten as:
\begin{equation}
\begin{aligned}
    &\int d \bm \kappa
    \tilde \N(\bm \nu,\Sigma )[-\bm \kappa]
    e^{2\pi i \tilde {\bm J}^T \bm \kappa}
    \tilde \N(\bm \mu',\Sigma')[\bm \kappa] = \\
    & 
    \quad \int d \bm \kappa
    e^{2\pi i \bm \kappa^T \bm \nu}
    e^{-2 \pi^2 \bm \kappa^T \Sigma \bm \kappa}
    e^{2\pi i \tilde {\bm J}^T \bm \kappa}
    e^{-2\pi i \bm \kappa^T \bm \mu'}
    e^{-2 \pi^2 \bm \kappa^T \Sigma' \bm \kappa}=\\
    & \qquad
    \int d \bm \kappa
    e^{-2\pi i \bm \kappa^T (\bm \mu'-\bm \nu-\tilde {\bm J})}
    e^{-2 \pi^2 \bm \kappa^T (\Sigma+\Sigma') \bm \kappa}.
\end{aligned}
\end{equation}
Computing the integral, this last equation can be cast as:
\begin{equation}
\begin{aligned}
    &\int d \bm \kappa
    \tilde \N( \bm \nu,\Sigma )[-\bm \kappa]
    e^{2\pi i \tilde {\bm J} ^T \bm \kappa}
    \tilde \N(\bm \mu',\Sigma')[\bm \kappa] = \\
    & \quad
    \frac{1}{\sqrt{(2\pi)^{M} \det(\Sigma + \Sigma ')}}
    \exp[{-\frac{1}{2}\bm \zeta^T (\Sigma + \Sigma ')^{-1} \bm \zeta }], 
\end{aligned}
\end{equation}
where we defined the quantity $\bm \zeta = \bm \zeta(\bm J,\tilde {\bm J})$ as: 
\begin{equation}
 \bm \zeta(\bm J,\tilde {\bm J}) := \bm \mu' - \bm \mu - \bm J^T \Sigma - \tilde {\bm J}.
\end{equation}
Oftentimes, as mentioned in the main text, it is convenient to work with complex Gaussians. This assumption, in the Variational Multi-Gaussian (VMG) ansatz, makes us be able to capture negative features of the Wigner function such as interference phenomena.
In the present work, this is achieved by considering an ansatz Wigner function given by the sum of the real parts of a set of complex Gaussians (cf. Eq.~\eqref{eq:vmg}), whose centers $\bm\mu$ are allowed to also have an imaginary part: $\bm\mu= \bm\alpha +i \bm\beta$. 
Effectively, the imaginary components of the centers act by re-modulating the Gaussian by a sinusoidal component, thus allowing negative values. 
The choice of taking the real parts of complex Gaussians is not arbitrary.  Indeed, while the real part of a complex Gaussian retains the normalization condition of a proper quasiprobability density function, the norm of the imaginary part is instead zero. 
In this case the generalized Gaussian moments can be computed as derivatives of a generating function $\mathcal{Z}[\bm J, \tilde{\bm J}]$: 
\begin{equation}
\begin{aligned}
    &\llangle
    \xi_{i_1}...\xi_{i_l}
    \partial_{\xi_{j_1}}...\partial_{\xi_{j_k}}
    \rrangle_{(\bm  \mu,\Sigma ),(\bm \mu',\Sigma')}=\\
    &\partial_{J_{i_1}}...\partial_{J_{i_l}}\partial_{\tilde J_{i_1}}...\partial_{\tilde J_{i_k}}\mathcal{Z}[\bm J, \tilde{\bm J}].
\end{aligned}
\end{equation}
Specifically, $\mathcal{Z}[\bm J, \tilde{\bm J}]$ has the form:
\begin{equation}\label{eq:ZJJ}
\begin{aligned}
    \mathcal{Z}[\bm J,\tilde {\bm J}]=
   \frac{\frac{1}{2} e^{\frac{1}{2}\bm J^T \Sigma \bm J+ \bm J^T \bm \alpha}}{\sqrt{(2\pi)^{M} \det(\Sigma + \Sigma ')}} \mathcal{Z}_1
   (\mathcal{M}_-\mathcal C _- + \mathcal{M}_+\mathcal C _+),
\end{aligned}
\end{equation}
where the function $\mathcal{Z}_{1}$ is defined as:
\begin{equation}
    \mathcal{Z}_{1}
    =
    \exp[-\frac{1}{2}
    (\bm \alpha'-\bm \alpha - \bm J^T \Sigma - \tilde {\bm J})^T
   (\Sigma + \Sigma')^{-1}
   (\bm \alpha'-\bm \alpha - \bm J \Sigma - \tilde {\bm J})
   ],
\end{equation}
while the functions $\mathcal{M}_-$ and $\mathcal{M}_+$ are:
\begin{equation}
\begin{aligned}
    &\mathcal{M}_-
    =
    \exp[\frac{1}{2}
    (\bm \beta'-\bm \beta )^T
   (\Sigma + \Sigma')^{-1}
   (\bm \beta'-\bm \beta )],
   \\
   &\mathcal{M}_+
    =
    \exp[\frac{1}{2}
    (\bm \beta'+\bm \beta )^T
   (\Sigma + \Sigma')^{-1}
   (\bm \beta'+\bm \beta )].
\end{aligned}
\end{equation}
Finally, the functions $\mathcal{C}_-$ and $\mathcal{C}_+$ are defined as:
\begin{equation}
\begin{aligned}
    &\mathcal{C}_-&
    = \cos \left[\bm J^T \bm \beta- (\bm \beta ' - \bm \beta)^T(\Sigma + \Sigma')^{-1}(\bm \alpha'-\bm \alpha -\bm J ^T\Sigma - \tilde {\bm J})\right],\\
    &\mathcal{C}_+&
    = \cos \left[\bm J^T \bm \beta +(\bm \beta ' + \bm \beta)^T (\Sigma + \Sigma')^{-1}(\bm \alpha'-\bm \alpha - \bm J^T \Sigma - \tilde{\bm J})\right].
\end{aligned}
\end{equation}

\fra{The time evolution of the variational parameters is obtained by automatically deriving this expression.
Specifically, we differentiate it in $\bm J$ and $\tilde{\bm J}$ with Taylor mode differentiation to obtain the various terms in the Liouvillian $\mathcal{L}$, and thus the Liouvillian gradient $\bm V$.
When evaluated at vanishing currents, it can be used to obtain the quantum geometric tensor $T$ by computing its Hessian with respect to the parameters $\bm \theta$.}

\section{Phase space formulation of quantum mechanics}
\label{appendix:phase_space}
In the phase-space formulation of quantum mechanics, there exists a one-to-one correspondence between operators and functions on position-momentum coordinates. 
For the sake of simplicity, we present the formalism for a single bosonic mode; the generalization to
M modes follows by tensor product. 
For any quantum operator $\hat O(\hat q, \hat p)$ we can define its associated Weyl symbol $O_W(q,p)$ as \cite{polkovnikov2010phase}:
\begin{equation}\label{eq:Weyl_symbol definition}
O_W(q, p)=\int d \jac{y}\left\langle q-\frac{\jac{y}}{2}\right| \hat{O}(\hat{q}, \hat{p})\left|q+\frac{\jac{y}}{2}\right\rangle \mathrm{e}^{i p \jac{y} / \hbar} .
\end{equation}
The Wigner function, in turn, is defined from the density matrix $\hat \rho$ as: 
\begin{equation}
W(q,p) = \frac{1}{\jac{2 \pi \hbar} }\int d y \  \hat\rho(q-y / 2, q+y / 2) \mathrm{e}^{i p y / \hbar} .
\end{equation}
Weyl symbols are functions of great importance in the phase-space formulation of quantum mechanics. The expectation value of any operator is given by the average of the corresponding Weyl symbol over the Wigner function: 
\begin{equation}
\begin{aligned}
\langle\hat{O}(\hat{q}, \hat{p})\rangle \equiv \operatorname{Tr}[\hat{\rho} \hat{O}(&\hat{q}, \hat{p})]=\int d q d p W(q, p) O_W(q, p).
\end{aligned}
\end{equation}

In order to introduce quantum dynamics in the phase-space formalism \cite{Vogel1989_dynamics}, we first need to introduce Moyal brackets $\{\circ , \circ \}_{MB}$ and the star product $\star$. We can define the star product between the Weyl symbol of an operator and the Wigner function as follows: 
\begin{equation}
\begin{aligned}
& H_{W} \star W=H_W\left(q+\frac{i \hbar}{2} {\partial_p}, p-\frac{i \hbar}{2} {\partial_q}\right) W, \\
& W \star H_{W}=H_W\left(q-\frac{i \hbar}{2} {\partial_p}, p+\frac{i \hbar}{2} {\partial_q}\right) W.
\end{aligned}
\end{equation}
Moyal bracket's $\{\circ , \circ \}_{MB}$ definition comes directly from the star product: 
\begin{equation}
\{H, W\}_{MB} = -\frac{i}{\hbar} (H_W\star W - W\star H_W). 
\end{equation}
The Lindbald master equation in phase space can be reformulated as follows: 
\begin{equation}\label{eq:Wig_phase_space_dynamics}
    \frac{\partial W(q,p)}{\partial t} = \{H_W,W\}_{\textit{MB}} + \sum_j \gamma_j {\mathcal D_W}_j \left[W\right ] , 
\end{equation}
where $H_W$ is the Weyl symbol of the Hamiltonian and $\{\circ,\circ\}_{\textit{MB}}$ represents the Moyal bracket. 
Moreover the action of the phase space super-operators ${\mathcal D_W}_j$ in the phase space, whose Weyl symbol for the jump operator is ${\Gamma_W}_j$, reads:
\begin{equation}\label{disspiation_Wigner}
\begin{aligned}
\mathcal{D}_j {W} &=
\sum_k {{\Gamma_W}_{jk}} \star {W} \star {\Gamma_W}_{jk}^{\dagger} \\
-\frac{1}{2}
&\left(
({{\Gamma_W}_{jk}}^{\dagger} {\Gamma_W}_{jk}) \star 
{W} + W \star ({\Gamma_W}_{jk}^{\dagger} {\Gamma_W}_{jk})
\right).
\end{aligned}
\end{equation}
By a quick comparison with the Lindblad master equation for the density matrix (Eq.~\eqref{eq:Lindblad}), it is possible to notice how the Moyal brackets replace the commutator, the Weyl symbols take the place of the operators, and the Wigner function fill in for the density matrix $\hat \rho$. 
\subsection{Lindbladian in phase-space}
\label{Bose-hubbardPhase}
Let us now consider a $M$-modes (sites) driven dissipative Bose-Hubbard Hamiltonian (Eq.~\eqref{eq:driven-diss-Bose-Hubb}) which we also report here for the sake of simplicity: 
\begin{equation}
\begin{aligned}
\hat{H} =\sum_{j=1}^{N_G}\Big[ -\Delta \hat{a}_j^{\dagger} \hat{a}_j+\frac{U}{2} \hat{a}_j^{\dagger 2} \hat{a}_j^2+\frac{G}{2} \hat{a}_j^{\dagger 2}+\frac{G^*}{2} \hat{a}_j^2 \Big] 
-\sum_{\left\langle j, j^{\prime}\right\rangle} \frac{J}{z}\left(\hat{a}_j^{\dagger} \hat{a}_{j^{\prime}}+\hat{a}_{j^{\prime}}^{\dagger} \hat{a}_j\right) .
\end{aligned}
\end{equation}
For clarity, we will denote by $\mathcal{L}_G W$ the explicit result of Eq.~\eqref{eq:Wig_phase_space_dynamics} comprising only the terms proportional to the double-boson driving amplitude $G$. We use an analogous notation for all the other terms in the Hamiltonian. 
By explicitly computing the terms in the Hamiltonian term of the Lindblad master equation Eq.~\eqref{eq:Wig_phase_space_dynamics} one finds that the expression for $\mathcal L_G$ reads as: 
\begin{equation}
\begin{aligned}
    \mathcal{L}_G W &= \sum_{j=1}^{M} \Big[ \text{Re}[ G] \left( p_j \frac{\partial W}{\partial q_j} + q_j \frac{\partial W}{\partial p_j} \right) 
    - \text{Im}[G] \left( q_j \frac{\partial W}{\partial q_j} - p_j \frac{\partial W}{\partial p_j} \right)\Big],
\end{aligned}
\end{equation}
while that for the detuning term $\mathcal L_\Delta$ is:
\begin{equation}
    \mathcal{L}_\Delta W = \Delta \sum_{j=1}^{M} \Big [   p_j \frac{\partial W}{\partial q_j} - q_j \frac{\partial W}{\partial p_j}  \Big],
\end{equation}
that for the Kerr non-linearity term $\mathcal L_U$ is: 
\begin{equation}
\begin{aligned}
    \mathcal{L}_U W = \frac{U}{2} \sum_{j=1}^{M} \Big[ 
    & (q_j^2 + p_j^2) 
    \left( q_j \frac{\partial W}{\partial p_j} - p_j \frac{\partial W}{\partial q_j} \right) 
     + 2 \left( p_j \frac{\partial W}{\partial q_j} - q_j \frac{\partial W}{\partial p_j} \right) \\
    & + \frac{1}{4} \Big( 
        p_j \frac{\partial^3 W}{\partial q_j^3}
        - q_j \frac{\partial^3 W}{\partial p_j^3} 
         - q_j \frac{\partial^3 W}{\partial p_j \partial q_j^2}
     + p_j \frac{\partial^3 W}{\partial q_j \partial p_j^2} 
    \Big) 
    \Big],
\end{aligned}
\end{equation}
and that for the nearest neighbor hopping term $\mathcal L_J$ is:
\begin{equation}
\begin{aligned}
    \mathcal{L}_J W =  
    -\frac{J}{z} \sum_{\langle j,j'\rangle } \Big[ q_j \frac{\partial}{\partial p_{j'} } + q_{j'}\frac{\partial}{\partial p_j} - p_j \frac{\partial}{\partial q_{j'}} - p_{j'} \frac{\partial}{\partial q_j}\Big]W.
\end{aligned}
\end{equation}
In a similar way one can also compute the terms corresponding to the single boson dissipation: 
\begin{equation}
\begin{aligned}
    \mathcal{L}_{\gamma} W = \frac{\gamma}{2} \sum_{j=1}^{M} \Big[  q_j \frac{\partial W}{\partial q_j} + p_j \frac{\partial W}{\partial p_j}  
    + 2 W + \frac{1}{2} \frac{\partial^2 W}{\partial q_j^2} + \frac{1}{2} \frac{\partial^2 W}{\partial p_j^2} \Big],
\end{aligned}
\end{equation}
and those related to the two-boson jump operators:
\begin{equation}
\begin{aligned}
    \mathcal{L}_\eta W &=  \frac{\eta}{2}   \sum_{j=1}^{M} \Big[ \Big( (q_j^3 + q_j p_j^2) \frac{\partial W}{\partial q_j}
     + (p_j^3 + q_j^2 p_j) \frac{\partial W}{\partial p_j} \Big) 
    + 2 \eta \left( (q_j^2 + p_j^2) W \right) + \eta \left( q_j \frac{\partial W}{\partial q_j} + p_j \frac{\partial W}{\partial p_j}\right) \\ 
    &+ \frac{\eta}{2} \left( (q_j^2 + p_j^2) \left( \frac{\partial^2 W}{\partial q_j^2} + \frac{\partial^2 W}{\partial p_j^2} \right) \right)  
    + \frac{\eta}{8} \big( q_j \frac{\partial^3 W}{\partial q_j^3} + p_j \frac{\partial^3 W}{\partial p_j^3} + q_j \frac{\partial}{\partial p_j} \big( \frac{\partial^2 W}{\partial p_j^2} \big) 
     + p_j \frac{\partial}{\partial p_j} \big( \frac{\partial^2 W}{\partial q_j^2} \big) \big) \Big].
\end{aligned}
\end{equation}
\fra{
Finally, for the non-local losses appearing in Sec. \ref{sec:Kerr-nonlocal}, the action of the differential operator $\mathcal{L}_{\kappa}W$ on the Wigner function, corresponding to the non-local jump operators $\hat \Gamma_\kappa=\sqrt \kappa[\hat a_j+\hat a_{j+1}]$ is given by:
\begin{equation}
    \label{eq:non-localL}
    \begin{split}
    &\mathcal{L}_{\kappa}W=
    \kappa\sum_{j=1}^3
    \Big[2W+
    \\
    &
    \left( 
    p_j\frac{\partial W}{\partial p_j }+
    q_j\frac{\partial W}{\partial q_j }
    \right)+
    \frac{1}{2}\left( 
    \frac{\partial^2 W}{\partial p_j ^2}+
    \frac{\partial^2 W}{\partial q_j ^2}
    \right)\Big]+
    \\
    &
    \kappa\sum_{j=1}^2
    \left[
    \frac{1}{2}\left(
    \frac{\partial^2W}
    {\partial p_j\partial p_{j+1}}+
    \frac{\partial^2W}
    {\partial q_j\partial q_{j+1}}
    \right)
    +
    p_j\frac{\partial W}{\partial p_{j+1} }+
    q_j\frac{\partial W}{\partial q_{j+1} }
    \right]
    \end{split}
\end{equation}
As can be seen, the terms in these expressions consist of monomials in $p_i$ and $q_i$ multiplied by derivatives with respect to these variables. When applied to the Wigner functions, their action can be equivalently expressed in terms of derivatives of the generating function $\mathcal Z[\bm J, \tilde{\bm J}]$, cf. Eq. \eqref{eq:ZJJ}. For example, a purely local term like $q_j\frac{\partial W}{\partial  p_j}$ can be computed as
\begin{equation}
    q_j\frac{\partial W}{\partial  p_j}=\partial_{J_{q_j}}\partial_{\tilde J_{p_j}} \mathcal{Z}[\bm J=\bm 0, \tilde{\bm J}=\bm 0],
\end{equation}
while for a non-local term such as $\frac{\partial^2 W}{\partial  p_i p_{i+1}}$, one obtains
\begin{equation}
    \frac{\partial^2 W}{\partial  p_i p_{i+1}}=\partial_{\tilde J_{p_i}} \partial_{\tilde J_{p_{i+1}}}\mathcal{Z}[\bm J=\bm 0, \tilde{\bm J}=\bm 0].
\end{equation}
In particular, the derivation is never explicitly in terms of the phase space variables $p_i$ or $q_i$.
}
\subsection{Parity Weyl symbol}
In this subsection, we calculate the Weyl symbol corresponding to the boson number parity operator $\hat \Pi = \exp (i\pi \sum_i\hat a _i^\dagger \hat a_i)$. 

The parity operator $\hat {\Pi}$ acts on the position and momentum basis as \cite{Haroche:993568}: 
\begin{equation}
\hat {\Pi}|q\rangle=|-q\rangle ; \quad \hat {\Pi}|p\rangle=|-p\rangle. 
\end{equation}
The Weyl symbol of the parity operator $\hat \Pi$ of a $1$-mode bosonic system can be computed directly from the definition Eq.~\eqref{eq:Weyl_symbol definition}:
\begin{equation}
\begin{aligned}
    \Pi_W(x, &p) = \int d \xi\left\langle x-\frac{\xi}{2}|\hat {\Pi}| x+\frac{\xi}{2}\right\rangle \mathrm{e}^{i p \xi / \hbar} 
    = \int d \xi \ \delta(2x) \frac{2 \pi}{2 \pi} e^{i p \xi / \hbar} \\
    &= \frac{\delta(x)}{2} \delta(p) 2\pi = \pi \delta(x) \delta(p).  
\end{aligned}
\end{equation}
In the phase-space formalism, the expectation value of the parity operator on the whole system is thus simply given by $\pi$ times the value of the Wigner function at the center of the phase space. Analogously, for a system of $n$-bosonic sites, we have found that it is given by: 
\begin{equation}
    \Pi_W(\boldsymbol q, \boldsymbol p) = \pi^n \delta^{(n)}(\bm q)\delta^{(n)}(\bm p), 
\end{equation}
which is the simple generalization to an $n-$dimensional space.
\end{widetext}
\section{Squeezed Coherent states}\label{sec:squeeze}
In a Gaussian function [Eq.~\eqref{eq:Gaussian}], the covariance matrix $\Sigma$ is of profound importance, as it determines the extent to which the function is squeezed along one direction relative to another. 
We have seen in the previous paragraph that the parity expectation value is determined by the value of the Wigner function in the origin of the phase-space. 
If we now imagine to have a Gaussian centered in zero and to arbitrarily reduce its covariance matrix -which controls the broadness of the Gaussian- its peak in zero will become higher and higher. Once the peak of the Gaussian overcomes the value of $1/\pi$, the parity expectation value would exceed one, thus violating the Heisenberg uncertainty principle. 
In this paragraph we introduce all the key elements and the basic notions to parametrize a physical covariance matrix for Gaussian quantum states. 

Let us consider an $M$-modes bosonic system with annihilation and creation operators obeying the standard commutation relations: 
\begin{equation}
\begin{aligned}
& {\left[\hat{a}_j, \hat{a}_k^{\dagger}\right]=\delta_{j k},} \\
& {\left[\hat{a}_j, \hat{a}_k\right]=\left[\hat{a}_j^{\dagger}, \hat{a}_k^{\dagger}\right]=0 .}
\end{aligned}
\end{equation}
It is usually convenient to introduce the Hermitian real quadrature operators:
\begin{equation}
\hat{q}_k=\frac{1}{\sqrt{2}}\left(\hat{a}_k+\hat{a}_k^{\dagger}\right), \quad \hat{p}_k=\frac{1}{i \sqrt{2}}\left(\hat{a}_k-\hat{a}_k^{\dagger}\right),
\end{equation}
which satisfy: 
\begin{equation}
\begin{aligned}
& {\left[\hat{q}_j, \hat{p}_k\right]=i \delta_{j k},} \\
& {\left[\hat{q}_j, \hat{q}_k\right]=\left[\hat{p}_j, \hat{p}_k\right]=0 .}
\end{aligned}
\end{equation}
If we arrange these real quadrature into the following vector \cite{Simon1994}:
\begin{equation}
\hat{\boldsymbol{\xi}}=\left(\begin{array}{c}
\hat{q}_1 \\
\hat{p}_1 \\
\vdots \\
\hat{q}_M \\
\hat{p}_M
\end{array}\right) \text {, }
\end{equation}
the canonical commutation relation can thus be rewritten as: 
\begin{equation}
\left[\hat{\boldsymbol{\xi}}_k, \hat{\boldsymbol{\xi}}_l\right]=i {\Omega}_{k l}.
\end{equation}
Here $\Omega$ is the symplectic matrix defined as
\begin{equation}
{\Omega}=\bigoplus_{k=1}^{2 M}\left(\begin{array}{cc}
0 & 1 \\
-1 & 0
\end{array}\right)=\mathbb{1}_M\otimes\left(\begin{array}{cc}
0 & 1 \\
-1 & 0
\end{array}\right).
\end{equation}
The covariance matrix of a generic pure Gaussian state can be written as: 
\begin{equation}
\Sigma_{k l}=\frac{1}{2}\left\langle\left\{\hat{\boldsymbol{\xi}}_k, \hat{\boldsymbol{\xi}}_l\right\}\right\rangle-\left\langle\hat{\boldsymbol{\xi}}_k\right\rangle\left\langle\hat{\boldsymbol{\xi}}_l\right\rangle . 
\label{eq:Sigma_Gauss}
\end{equation}
The covariance matrix $\Sigma$, besides being real, symmetric, and positive-definite, must also satisfy the quantum uncertainty principle, which requires the matrix $\Sigma + \frac{i}{2}\Omega$ to be positive semi-definite \cite{Simon1994}.

In general, checking the positive semi-definiteness of a generic covariance matrix $\Sigma$ is a nontrivial task. 
In order to overcome this issue we
construct a physical covariance matrix by applying squeezing and displacement operators \cite{Walls2008}. 
Namely, a squeezed coherent state is defined as
\begin{equation}
    \ket{\alpha,\zeta} =    \hat D(\alpha) \hat S(\zeta)\ket{0},
\end{equation}
where $\ket0$ is the vacuum state,  while $\hat D$ and $\hat S$ are respectively the displacement and the squeezing operator:  
\begin{equation}
\begin{aligned}
&\hat{D}(\bm \alpha)=\prod_{j=1}^M\exp \left(
\alpha_j \hat{a_j}^{\dagger}-\alpha_j^* \hat{a}_j\right) , \\
&\hat{S}(\bm \zeta)=\prod_{j=1}^M\exp \left[\frac{1}{2}
\left(\zeta_j^* \hat{a}^2_j-\zeta_j\hat{a}_j^{\dagger 2}\right)\right]. 
\end{aligned}
\end{equation}
Both these operations are symplectic, i.e. $\Omega$ is left invariant under their action: $\hat S^T \Omega \hat S = \Omega$ and $\hat D^T \Omega \hat D = \Omega$. This implies that any state whose covariance matrix $\Sigma$ is obtained through displacement and squeezing operations respects the condition for $\Sigma + \frac{i}{2}\Omega$ to be positive semi-definite \cite{Walschaers2021,Simon1994}: $\Sigma + \frac{i}{2}\Omega\geq 0$. 
Specifically, the covariance matrix of the single-mode vacuum state ($\Sigma_{VAC} =\mathbb 1_2 /2$) is transformed by the squeezing operator $\hat S(\zeta = re^{i\phi})$ as \cite{Bourassa2021}:
\begin{equation}
\Sigma(\zeta) =\frac{1}{2} \cosh (2 r) \mathbb{1} _2 -\frac{1}{2} \sinh (2 r) \boldsymbol{S}_\phi,
\end{equation}
where 
\begin{equation}
\boldsymbol{S}_\phi=\left(\begin{array}{cc}
\cos (\phi) & \sin (\phi) \\
\sin (\phi) & -\cos (\phi)
\end{array}\right) .
\label{eq:sigma_param}
\end{equation}
The displacement operator leaves the covariance matrix unchanged.

\section{Variational Multi-Gaussian parametrization}\label{app:parametrization}
In this section we expose how we parametrize the Variational Multi-Gaussian (VMG) ansatz for the Wigner function and we discuss how the number of variational parameters scales with the number of Gaussians ($N_G$) and the number of bosonic modes ($M$). \\ 
The vector $\bm \theta$ includes all the ansatz variational parameters, namely the normalization coefficients $c_i$, the centers $\bm {\mu}_i$ and the squeezing parameters $\zeta_i$, where the index $i$ ranges from $1$ to $N_{\rm G}$.

Since the Liouvillian terms considered in this work are invariant under the transformation $\hat a_j \xrightarrow[]{} - \hat a_j$ and $\hat a_j^\dagger \xrightarrow[]{} - \hat a_j^\dagger $, the quantum master equation preserves the phase-space parity of the Wigner function ($W(\bm q, \bm p) = W(-\bm q,- \bm p)$). Since the steady state is also expected to be an even phase-space function, we consider a variational ansatz given by: 
\begin{equation}
\begin{aligned}
    W(\boldsymbol \xi; \boldsymbol{\theta}) &\equiv \sum_{i=1}^{N_{\rm G}} {\rm Re}\Big[G\big(\boldsymbol \xi; \boldsymbol\theta_i =(c_i,\bm \mu_i, \Sigma_i)\big)
    + \\
    &G\big(\boldsymbol \xi; \boldsymbol\theta_i=(c_i,- \bm \mu_i, \Sigma_i)\big)\Big],
\end{aligned}
\end{equation}
which is explicitly even. 
Moreover, the centers of the Gaussians are assumed to be complex: $\boldsymbol \mu =\boldsymbol \alpha + i \boldsymbol \beta$ in order to capture negative features. 
To parameterize the ansatz for $2N_G$ Gaussian functions, we need $N_G$ real parameters for the normalization coefficients, and $N_G   \times 4 M$ real parameters for the real and the imaginary parts of the centers. 

\jac{For the system described in Sec. \ref{sec:Kerr-nonlocal}, we parameterized the covariance matrix as $\Sigma = L^\dagger L$, where $L$ is a real matrix. This construction guarantees the positive semi-definiteness of $\Sigma$. We initialized $L$ in such a way that, at the initial time $t=0$, $\Sigma$ corresponds to the vacuum covariance matrix ($ \bm 1 /2$) perturbed by small numerical noise on the diagonal.
In this case we employed 64 pairs of Gaussian functions. For a $3$-mode system, in order to encode the $L$ matrix for each Gaussian, $36$ real parameters are needed. 
Since we employed a total number of $64$ pairs of Gaussian functions, a total of $3136$ real parameters are needed when also accounting for the centers $\bm \mu_i$ and the coefficients $c_i$. Within section \ref{sec:Kerr-nonlocal}, the coefficients $c_i$ are left free to assume both positive and negative values while keeping their total sum normalized to 1.}

In the case of \jac{the systems described in Sec. \ref{sec:single-kerr} and Sec. \ref{sec:bose-hubbard-steady}} we initialized the covariance matrices $\Sigma_i$ as a block diagonal matrices in which every block is determined by $M$-couples of parameters encoding $M$-single-mode squeezing operations on the vacuum state: $(\bm r$,$\bm \phi)$, where $\bm r = (r_1, ..., r_{M})$ and $\bm \phi = (\phi_1, ...,\phi_{M})$ \jac{while $M$ is the total number of bosonic modes of the system} [cf. Eq.~\eqref{eq:sigma_param}].  
While multi-mode correlations are still guaranteed by the presence of many Gaussian functions with different weights $c_i$, this parametrization scales linearly with the number of Gaussian functions. 
For the considered case of bi-Gaussians, in order to encode the covariance matrices of $2 N_{\rm G}$ Gaussians, we need $N_{G} \times {2M}$ real parameters. \\
Eventually, to fully encode $2 N_{\rm G}$ Gaussian functions we thus need a total number of $6 N_{\rm G} \times (M+1/6) $ variational parameters. For example, to compute the dynamics of a $144$ bosonic sites with $16$ Gaussians, we unraveled the dynamical evolution of $6920$ parameters. 
This approach scales linearly both in the number $N_G$ of Gaussians and in the number $M$ of bosonic modes considered. \\
\jac{Since we explicitly evolve the set of parameters $\bm \theta$, special care must be taken with respect to their initialization. \jac{With this kind of parametrization, accomodating as initial state a general mixtures of $N$ squeezed coherent states can be perfomed exactly since, in phase-space, it is represented exactly by a sum of $N$ real-centered Gaussians, whose covariance matrix is uniquely determined by the squeezing factors.} \\ 
In Sec. \ref{sec:single-kerr} and Sec. \ref{sec:bose-hubbard-steady}, we adopt two different initial sets. As the first state $W_{\rm init,\, 0}$ we adopt a slight perturbation of the vacuum state where the squeezing parameters $(\bm r, \bm \phi)$ are initialized at random near zero. 
Specifically, we choose them uniformly distributed in the interval $(0,10^{-4}]$. The real parts of the centers $\alpha_i$ is chosen  uniformly distributed between $0$ and $10^{-3}$ and the imaginary part $\beta_i$ between $10^{-3}$ and $10^{-5}$. 
The second initial state we consider, $W_{\rm init,\, 1}$,  amounts to a small perturbations of the vacuum plus a displacement. This state is characterized by a non negligible initial number of photons. In this case, $(\bm r, \bm \phi)$ are uniformly randomly initialized in the interval $(0,0.01]$ while $\beta_i$ between $1.4\times 10^{-4}$ and $1.6\times 10^{-5}$. The parameters $\alpha_i$ are instead initialized as uniformly distributed between $0.13$ and $0.14$. Unless explicitly stated, all the figures int the paper were obtained using the initial condition $W_{\rm init,\, 0}$. \\
The choice of initial state $W_{\rm init,\, 1}$ was driven by the fact that, for low driving amplitudes $G$, the steady state is close to the bosonic vacuum. Initializing the system in this region would suppress transient dynamics and make the numerical fit of the decay rate more difficult. Therefore, we initialized the system in a state ``distant'' from the vacuum to enhance the transient dynamics. Moreover, as the initial state, we introduced a small perturbation around the vacuum. Initializing all Gaussian functions in the ansatz exactly in the vacuum state would render the geometric tensor $M$ in Eq.~\eqref{eq:DiracFrank} degenerate, thereby preventing the subsequent evolution. }
\jac{\section{Regularized 
Equations of Motion}\label{app:regularization}}
\jac{The equations of motion governing the dynamics of the variational parameters, which we report here for clarity, involve the inversion of the quantum geometric tensor $T$:
\begin{equation}
    T \cdot \frac{d}{dt}\boldsymbol{\theta} = \boldsymbol{V}. \label{eq:DiracFrank}
\end{equation}
The numerical solution for $\dot{\boldsymbol{\theta}}$ may become unstable when the quantum geometric tensor is nearly singular. Numerical inversion therefore requires regularization; to achieve this, we add a small diagonal shift to the matrix $T$. The equations of motion thus become:
\begin{equation}
    \dot{\boldsymbol{\theta}} = (T[\boldsymbol{\theta}] + \lambda \bm{1})^{-1} \boldsymbol{V}[\boldsymbol{\theta}].
\end{equation}}
Typical values for the diagonal shifts implemented in the dynamics are $\lambda = 10^{-10}$ for the many-mode driven-dissipative Bose-Hubbard model and $\lambda = 10^{-12}$ for a single Kerr cavity. In particular, to solve the linear system involving $(T[\boldsymbol{\theta}] + \lambda \bm{1})$, we utilized Cholesky decomposition~\cite{Golub1996}. 
This regularization procedure is formally equivalent to an L2-regularized least-squares problem. Introducing the shift $\lambda$ corresponds to finding the update vector $\dot{\boldsymbol{\theta}}$ that minimizes a cost function with a penalty on the vector norm \cite{Medvidovic2023}:
\begin{equation}
    \dot{\boldsymbol{\theta}} = \underset{\boldsymbol{x} \in \mathbb{R}^P}{\mathrm{argmin}} \left\{ \left\| L \boldsymbol{x} - \boldsymbol{y} \right\|_2^2 + \lambda \left\| \boldsymbol{x} \right\|_2^2 \right\},
\end{equation}
where $L$ is the Cholesky factor of the quantum geometric tensor (such that $L^T L = T$), $\boldsymbol{y}$ satisfies $L^T \boldsymbol{y} = \boldsymbol{V}$, and $\|\cdot\|_2$ denotes the standard Euclidean 2-norm.
We remark that the diagonal shifts adopted here for regularization are significantly smaller than those typically used when inverting the quantum geometric tensor in cases where its estimate is obtained via Monte Carlo sampling, which are often on the order of $10^{-3} - 10^{-6}$~\cite{Vicentini_2022}. This is because, for the variational multi-Gaussian ansatz, the quantum geometric tensor is computed analytically and is therefore free of the noise due to stochastic sampling.\\

\section{Taylor-mode automatic differentiation}
\label{app:taylor-mode}

Evaluating the generalized Gaussian moments requires computing many high-order
mixed derivatives of the generating function $\mathcal{Z}[\bm{J},\tilde{\bm{J}}]$
with respect to the vanishing currents $\bm{J}$ and $\tilde{\bm{J}}$. In this
Appendix we briefly introduce \emph{Taylor-mode} automatic differentiation and
explain why it evaluates such derivatives at a computational cost that improves
factorially over standard forward-mode automatic differentiation as the
derivative order increases~\cite{GriewankUtkeWalther2000}.

Ordinary forward-mode automatic differentiation propagates, together with the
value of a function, its first derivative along a chosen direction, by applying
the chain rule to each elementary operation~\cite{Margossian2019}. A single pass
through the calculation thus returns both $f(\bm{x})$ and the directional
derivative $\nabla f(\bm{x})\cdot\bm{s}$. Higher-order derivatives are then
obtained by repeatedly applying the same procedure. Reconstructing a full
derivative tensor in this way is, however, wasteful, because a mixed derivative
does not depend on the order in which the differentiations are carried out. As a
simple example, a Hessian has $n^{2}$ entries $\partial_i\partial_j f$, but only
$n(n+1)/2$ of them are distinct, since
$\partial_i\partial_j f=\partial_j\partial_i f$. At order $d$ the same symmetry
is far more pronounced: the derivative $\partial_{i_1}\!\cdots\partial_{i_d}f$ is
invariant under any permutation of its indices, so the $d!$ orderings all yield
the same number. As a result, for a function of $n$ variables, repeated
forward-mode differentiation manipulates all $n^{d}$ ordered choices of which
variables to differentiate, even though only
$\binom{n+d-1}{d}\approx n^{d}/d!$ of them are truly distinct.

Taylor-mode automatic differentiation removes this redundancy by propagating,
for each intermediate quantity in the calculation, an entire truncated Taylor
expansion rather than only its first derivative. Along a direction $\bm{s}$ one
writes
\begin{equation}
\begin{aligned}
f(\bm{x}+t\,\bm{s})&=\sum_{m=0}^{d} f_{[m]}(\bm{x};\bm{s})\,t^{m}
+O\!\left(t^{d+1}\right),\\[2pt]
f_{[m]}(\bm{x};\bm{s})&=\frac{1}{m!}\,\frac{\mathrm{d}^{m}}{\mathrm{d}t^{m}}\,
f(\bm{x}+t\,\bm{s})\Big|_{t=0}.
\end{aligned}
\label{eq:taylor-def}
\end{equation}
The coefficient $f_{[m]}$ is simply the $m$-th directional derivative of $f$
along $\bm{s}$, up to the factor $1/m!$. Crucially, each coefficient packs the
order-$m$ derivatives along $\bm{s}$ into a single number, with the permutation
symmetry already built in.

Any function is ultimately built from a sequence of elementary
operations---additions, multiplications, and standard functions such as the
exponential or the logarithm---each acting on the results of the previous ones.
Standard automatic differentiation exploits this structure by propagating a
first derivative through the calculation via the chain rule; Taylor-mode follows
exactly the same principle, but propagates the whole list of Taylor coefficients
at once. All that is needed is a rule that, for each elementary operation, gives
the coefficients of its output from those of its inputs. When two quantities are
added, their coefficient lists are added term by term. When they are multiplied,
$w=uv$, the coefficients combine just as when two polynomials in $t$ are
multiplied: the coefficient of $t^{m}$ collects every product of a term
$u_{j}\,t^{j}$ with a term $v_{m-j}\,t^{m-j}$ whose powers add up to $m$,
\begin{equation}
w_{m}=\sum_{j=0}^{m}u_{j}\,v_{m-j}\qquad (w=uv).
\label{eq:cauchy}
\end{equation}
This is nothing but the Leibniz rule for the $m$-th derivative of a product, with
the factorials absorbed into the definition $f_{[m]}=f^{(m)}/m!$ of the
coefficients. The standard functions (exponential, logarithm, trigonometric
functions, division) obey equally simple recurrences, which follow from the
differential equations they satisfy. A single pass through the calculation
therefore delivers all coefficients up to degree $d$---and hence the directional
derivatives of every order $0,1,\dots,d$ along $\bm{s}$ at once---at a cost of at
most $O(d^{2})$ per elementary operation.

To reconstruct every mixed derivative up to order $d$, one propagates one such
Taylor expansion for each \emph{distinct} top-order derivative---exactly
$\binom{n+d-1}{d}$ of them---and combines the results through a stable,
division-free interpolation~\cite{GriewankUtkeWalther2000}. The number of
quantities that must be propagated is therefore
\begin{equation}
\binom{n+d-1}{d}\;\approx\;\frac{n^{d}}{d!},
\label{eq:count}
\end{equation}
in place of the $n^{d}$ ordered derivatives handled by repeated forward-mode
differentiation: the cost is thus reduced by a factor that grows factorially
with the derivative order $d$. This reorganization additionally trades the
irregular access patterns of symmetric tensors for a regular, contiguous memory
layout~\cite{GriewankUtkeWalther2000}, which is particularly well suited to the
GPU architectures employed in the present work.

In our case, $f$ is the generating function $\mathcal{Z}[\bm{J},\tilde{\bm{J}}]$
of Eq.~\eqref{eq:re_ggm}, and the differentiation variables are the currents
$\bm{J}$ and $\tilde{\bm{J}}$. Each Liouvillian term
$\mathcal{L}_{W,\mathbf{I}}\propto
\xi_{i_1}\!\cdots\xi_{i_l}\,\partial_{\xi_{j_1}}\!\cdots\partial_{\xi_{j_k}}$
corresponds to a mixed derivative of $\mathcal{Z}$ of order $l+k$, which is at
most four for Hamiltonians quartic in the bosonic operators and jump operators
at most quadratic. Taylor-mode automatic differentiation, as implemented in the
\texttt{jax.experimental.jet} module of JAX~\cite{Bettencourt2019}, evaluates all
these current derivatives in a single pass, whereas the outer derivatives with
respect to the variational parameters $\bm{\theta}$ are taken with standard
forward-mode automatic differentiation. It is this combination that makes the
evaluation of the Liouvillian gradient $\bm{V}$ and the quantum geometric tensor
$T$ efficient enough to enable the simulation of the large multi-mode systems
studied in this work.

\bibliography{biblio}
\end{document}